\documentclass[conference]{IEEEtran}
\ifCLASSOPTIONcompsoc
 \usepackage[caption=false,font=normalsize,labelfont=sf,textfont=sf]{subfig}
\else
 \usepackage[caption=false,font=footnotesize]{subfig}
\fi
\usepackage{cite}
\usepackage{ifpdf}
\usepackage{tabularx}
\usepackage{subcaption}
\usepackage{graphicx}
\usepackage{float}
\usepackage{array}
\usepackage{amsmath}    
\usepackage{amssymb}    
\usepackage{diagbox}
\usepackage{makecell}
\usepackage{algorithm}
\usepackage{algpseudocode}
\usepackage{multirow}
\usepackage{xcolor}
\usepackage{booktabs}
\usepackage{url}
\usepackage{makecell} 
\usepackage[colorinlistoftodos,prependcaption,textsize=scriptsize,textwidth=1.2cm]{todonotes}
\AtBeginDocument{%
  }
\renewcommand{\vec}[1]{\mathbf{#1}} %

\newcommand{\sect}{\textsection}

\begin{document}

%
\title{SVDefense: Effective Defense against \\ 
Gradient Inversion Attacks via \\ 
Singular Value Decomposition}

\makeatother
\author{\IEEEauthorblockN{Chenxiang Luo}
	\IEEEauthorblockA{City University of Hong Kong\\
		chenxiluo8-c@my.cityu.edu.hk}
    \and
	\IEEEauthorblockN{David K.Y. Yau}
	\IEEEauthorblockA{Singapore University of Technology and Design\\
		david\_yau@sutd.edu.sg}
    \and
	\IEEEauthorblockN{Qun Song*\thanks{*Corresponding author.}} 
	\IEEEauthorblockA{City University of Hong Kong\\
		qunsong@cityu.edu.hk}
        }


%


\IEEEoverridecommandlockouts
\makeatletter\def\@IEEEpubidpullup{6.5\baselineskip}\makeatother

\maketitle

\begin{abstract}
Federated learning (FL) enables collaborative model training without sharing raw data but is vulnerable to gradient inversion attacks (GIAs), where adversaries reconstruct private data from shared gradients. Existing defenses either incur impractical computational overhead for embedded platforms or fail to achieve privacy protection and good model utility at the same time. Moreover, many defenses can be easily bypassed by adaptive adversaries who have obtained the defense details. To address these limitations, we propose \textit{SVDefense}, a novel defense framework against GIAs that leverages the truncated Singular Value Decomposition (SVD) to obfuscate gradient updates. \textit{SVDefense} introduces three key innovations, 
a Self-Adaptive Energy Threshold that adapts 
to client vulnerability,
a Channel-Wise Weighted Approximation that selectively preserves essential gradient information for effective model training while enhancing privacy protection, 
and a Layer-Wise Weighted Aggregation for effective model aggregation under class imbalance. 
Our extensive evaluation shows that \textit{SVDefense} outperforms existing defenses across multiple applications, including image classification, human activity recognition, and keyword spotting, by offering robust privacy protection with minimal impact on model accuracy. Furthermore, \textit{SVDefense} is practical for deployment on various resource-constrained embedded platforms. We will make our code publicly available upon paper acceptance.
\end{abstract}


%
\IEEEpeerreviewmaketitle

\section{INTRODUCTION}

\begin{figure}[t!]
\centering
\includegraphics[width=.9\columnwidth]{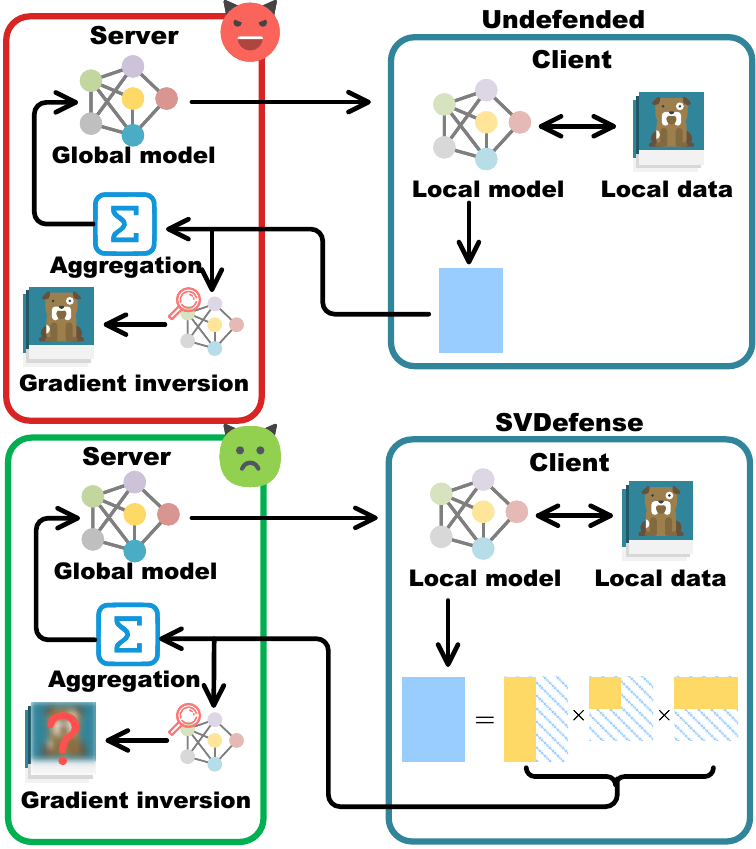}
\centering
\caption{{ An illustration of \textit{SVDefense}.} While the adversary may attempt to reconstruct user data using the gradients uploaded by an undefended client, our defense hinders the data reconstruction by uploading the gradients protected by \textit{SVDefense}. The blue and yellow blocks represent the original and truncated gradients in our defense, respectively.
}
\label{fig:intro-fig}
\end{figure}

Federated learning (FL) has emerged as a promising paradigm for collaborative model training that preserves user privacy in domains such as healthcare~\cite{dayan2021federated}, finance~\cite{ye2024openfedllm}, and social safety~\cite{meng2022improving}. In FL, distributed clients train models on their local datasets and only share model updates with a central server, eliminating the need to expose sensitive user data~\cite{mcmahan2017communication}.
However, recent research has revealed that FL systems remain vulnerable to gradient inversion attacks (GIAs)~\cite{zhu2019deep,geiping2020inverting,yue2023gradient}, where adversaries can reconstruct private training data by exploiting the gradients shared during client-server communication. These attacks have shown the ability to reconstruct private training data with high fidelity~\cite{li2022auditing,yue2023gradient,wu2023learning}, which raises substantial privacy risks in FL systems.


While various defense mechanisms against GIAs have been proposed, they face significant limitations in practical FL deployments. Encryption-based methods such as
secure multi-party computation (SMC)~\cite{bonawitz2017practical,lia2020privacy} and homomorphic encryption (HE)~\cite{kim2018efficient,zhang2020batchcrypt}
provide strong theoretical privacy guarantees but introduce prohibitive computational overhead for resource-constrained devices. Perturbation-based defenses counteract GIAs by
modifying local inputs~\cite{zhang2017mixup,gao2023automatic,huang2020instahide,wu2024concealing}, gradients~\cite{zhu2019deep,wei2021gradient,wang2022protect,wang2023more,zhang2025censor}, or training processes~\cite{scheliga2023dropout,scheliga2022precode,wan2023enhancing}. Defenses that perturb local inputs~\cite{zhang2017mixup,gao2023automatic,huang2020instahide,wu2024concealing} or add noise to local gradients,
such as differential privacy (DP)
and its variants~\cite{zhu2019deep,wei2021gradient,wang2022protect,wang2023more}, often fail to balance privacy protection and model utility~\cite{wang2023more,xue2024revisiting}. In our analysis, methods that perturb local gradients~\cite{zhang2025censor} and training
processes~\cite{scheliga2022precode,wan2023enhancing} can be bypassed by adaptive adversaries who have obtained details of the defense mechanisms. 
Pruning-based defenses~\cite{zhu2019deep,sun2021soteria,xue2024revisiting} that selectively remove gradient components to counteract GIAs can also be bypassed by adaptive attackers, as shown in our analysis. Unlike previous work~\cite{wu2023learning,balunovicbayesian} that considers less practical adaptive attacks relying on strong assumptions, we demonstrate the vulnerability of these defenses under GIAs augmented with practical adaptive operations.

This paper introduces \textit{SVDefense}, a novel defense framework against GIAs based on the truncated Singular Value Decomposition (SVD). 
Truncated SVD is a matrix factorization technique that approximates a matrix with the goal of effectively reducing its dimensionality while preserving important information.
{ Our design is motivated by key insights from the analysis of existing defenses under practical adaptive attacks, which suggests that affecting all the gradient components and doing so irreversibly are desirable properties that offer improved robustness against adaptive GIAs. Moreover, it is crucial for defense mechanisms to balance defense performance with model utility.}
Truncated SVD has the potential to counteract adaptive GIAs via gradient decomposition and truncation that irreversibly modifies the entire gradient space while preserving model utility. However, applying truncated SVD against GIAs presents practical challenges. Our preliminary study finds that clients with higher degrees of class imbalance are more vulnerable to attacks. Thus, more imbalanced clients should be given stronger protection from GIAs through lower energy thresholds $\mathcal{T}$ that reduce the information in truncated gradients for data reconstruction.
This strategy raises three key questions during the SVD truncation: 
(1) How to effectively quantify the varying degrees of class imbalance and adaptively adjust the energy thresholds $\mathcal{T}$ for different clients; (2) How to preserve information critical for model training while suppressing sensitive information leakage; and (3) How to effectively aggregate SVD truncated client updates and thereby improve global model utility under the class imbalance?

To address these challenges, \textit{SVDefense} introduces three key innovations. 
First, we observe that the distribution of the singular values obtained in SVD strongly correlates with the degree of class imbalance.
Hence, we propose a \textit{Self-Adaptive Energy Threshold} that adaptively adjusts the energy threshold for each client based on its singular value distribution, providing stronger protection to class-imbalanced clients who are more susceptible to GIAs.
Second, whereas lowering the energy threshold in SVD truncation provides stronger protection against GIAs, it also reduces the information needed for effective model training. To address this issue, we propose a \textit{Channel-Wise Weighted Approximation} that strategically assigns weights to gradients during the SVD truncation, which preserves gradients that are critical for model performance while suppressing potential sensitive information leakages, leading to better model accuracy and defense performance.
Third, non-IID local data distributions, such as class imbalance across the clients, lead to degraded global model accuracy due to client drift~\cite{wan2023enhancing}. 
Existing studies have not addressed how to effectively aggregate SVD truncated updates to improve the global model's utility under such heterogeneous data distributions. Our \textit{Layer-Wise Weighted Aggregation} addresses this gap by leveraging key correlation between singular value distributions and local class imbalance. By strategically assigning layer-wise aggregation weights to client updates based on their singular value distributions, we effectively improve the global model's accuracy under class-imbalanced data.
Together, the three components enable \textit{SVDefense} to achieve strong privacy protection while maintaining model utility across diverse FL scenarios. An illustration of how \textit{SVDefense} works is shown in Fig.~\ref{fig:intro-fig}.

Our extensive evaluation on the EMNIST~\cite{cohen2017emnist}, CIFAR-10~\cite{krizhevsky73cifar}, HAR~\cite{anguita2013public}, and Google Speech Commands~\cite{warden2018speech} datasets demonstrates that \textit{SVDefense} achieves superior performance in both model accuracy and defense effectiveness compared with various representative defenses. 
Moreover, we implement a real-world FL testbed on various embedded platforms, including Raspberry Pi, Nvidia Jetson Orin Nano, and Nvidia Jetson TX2, to validate the practicality of \textit{SVDefense}. Our experiments show that \textit{SVDefense} has high computational efficiency on various resource-constrained embedded platforms and significantly reduces communication cost. 

Our contributions are summarized as follows.
\begin{itemize}
\item We systematically analyze various representative defenses against GIAs, demonstrating their vulnerability to practical adaptive attacks. We derive key insights from our analysis to identify truncated SVD as a promising technique for leveraging these insights. 

\item We develop \textit{SVDefense}, a novel defense framework against adaptive GIAs based on truncated SVD. To the best of our knowledge, ours is the first comprehensive solution to address practical challenges of defending against GIAs under non-IID data distributions caused by class imbalance across FL clients. 
We introduce the Self-Adaptive Energy Threshold to adapt the privacy protection for clients based on their varying degrees of vulnerability to GIAs caused by their respective levels of class imbalance, the Channel-Wise Weighted Approximation to enhance both accuracy and defense performance, and the Layer-Wise Weighted Aggregation for effective aggregation of SVD truncated client updates to improve global model accuracy under the class imbalance.
\item Our extensive evaluation on various datasets 
demonstrates that \textit{SVDefense} outperforms existing defenses in both model accuracy and defense effectiveness. We also implement our solution in a real-world FL testbed using various embedded platforms.
Experimental results show that \textit{SVDefense} achieves practical computational cost and significantly reduces communication cost.
\end{itemize}

\section{BACKGROUND AND RELATED WORK}
\subsection{Gradient Inversion Attacks}

Existing GIAs 
can be categorized into optimization-based and GAN-based attacks. 

\noindent \textbf{Optimization-based Attacks:} Deep Leakage from Gradients (DLG)~\cite{zhu2019deep} is the first work that demonstrates the feasibility of reconstructing local data and corresponding labels from the shared gradients by iteratively optimizing dummy inputs to match the shared gradients using L-BFGS optimization.
Improved DLG (iDLG)~\cite{zhao2020idlg} enhances data reconstruction effectiveness by extracting ground-truth labels from gradient signs.
Inverting Gradients (IG)~\cite{geiping2020inverting} attack improves over the early studies of DLG and iDLG by employing the Adam optimizer to stabilize convergence and introduces cosine similarity as a more effective gradient matching objective.
GradInversion~\cite{yin2021see} can reconstruct high-fidelity input image batches using the gradient matching objective with fidelity regularization and group consistency regularization terms to improve reconstruction quality. However, its reliance on input batch normalization statistics makes it impractical in typical FL settings. 

\noindent \textbf{GAN-based Attacks:}
Generative adversarial networks (GANs)~\cite{goodfellow2020generative} are generative models capable of capturing the probability distribution of images from the training set.
Recent advances in GIAs leverage GANs as \textit{image} priors to compensate for information loss and enhance reconstruction quality. 
GIAS~\cite{jeon2021gradient} alternately optimizes latent vectors and generator parameters to improve image reconstruction fidelity.
GGL~\cite{li2022auditing} employs pre-trained GANs as priors to constrain the image reconstruction. GIFD~\cite{fang2023gifd} sequentially explores the latent space and intermediate features of the generator under an $l_1$-ball constraint to address limitations of expressiveness and generalization in pre-trained GANs.
A recent strong attack ROG~\cite{yue2023gradient} encodes raw images into low-dimensional representations to improve attack optimization efficiency, followed by GAN-based post-processing to enhance image reconstruction quality.

\noindent \textbf{Adaptive Attacks:}
Research has shown that an \textit{adaptive adversary}~\cite{tramer2020adaptive,wu2023learning,jiang2022primask} with knowledge of the defense (e.g., an honest-but-curious aggregator in FL who has legitimate access to the defense mechanism) can design targeted attacks against it. 
By formulating GIAs within a Bayesian framework, the work~\cite{balunovicbayesian} demonstrates how a Bayes optimal adversary can break several heuristic defenses. However, the theoretical analysis is based on specific neural network architectures.
Learning To Invert (LTI)~\cite{wu2023learning} trains a gradient inversion model to invert gradients protected by defenses including sign compression, gradient pruning, and gradient perturbation. However, LTI assumes that the adversary has access to the private data distribution.
The work~\cite{scheliga2023dropout} shows how Dropout's effectiveness as a defense against GIAs can be mitigated by modeling dropout-induced stochasticity during attack optimization.
While prior efforts~\cite{balunovicbayesian, wu2023learning} shed light on how adaptive adversaries can circumvent certain defenses, their findings are limited by strong assumptions about specific model architectures or attacker's capabilities.
In this paper, we extensively investigate the vulnerability of various representative defenses under realistic adaptive adversaries by augmenting existing GIAs with practical adaptive operations and further propose a novel defense framework.


\subsection{Gradient Inversion Defenses}
Existing defenses against GIAs can be categorized into encryption-based, perturbation-based, pruning-based, { and compression-based methods}.

\noindent \textbf{Encryption-based} defenses employ cryptographic techniques to protect client updates in FL.
SMC protocols~\cite{bonawitz2017practical, lia2020privacy} enable secure aggregation of client updates without revealing individual contributions.
Some studies~\cite{zhang2020batchcrypt, kim2018efficient} leverage HE to perform arbitrary computations on encrypted gradients.
Several efforts~\cite{sebert2023combining,chen2022poisson} combine DP with HE or SMC to provide formal privacy guarantees while allowing encrypted gradient aggregation.
Although these defenses offer theoretical privacy guarantees, they introduce significant computational, communication, and storage overhead
and often necessitate modifications to FL architectures, making them less practical.

\noindent \textbf{Perturbation-based}
defenses can be further divided into three sub-categories. 
\textit{Input perturbation} modifies the local training data. Approaches include creating composite images through linear combinations~\cite{zhang2017mixup, huang2020instahide}, applying strategic data augmentation~\cite{gao2021privacy,gao2023automatic}, and synthesizing visually distinct concealed samples to mimic sensitive data at the gradient level~\cite{wu2024concealing}. However, these methods often compromise classification performance, provide insufficient protection, or incur high computational overhead~\cite{xue2024revisiting}. 
\textit{Gradient perturbation} modifies the local gradients. 
Early defenses~\cite{zhu2019deep} leverage DP by adding Gaussian and Laplace noises to the gradients.
The work~\cite{wei2021gradient} applies per-example gradient clipping and DP noise injection during local training. 
The work~\cite{wang2022protect} adds layer-wise random perturbations to gradients based on information leakage risk.
Outpost~\cite{wang2023more} adaptively adds Gaussian noise combined with gradient pruning during each local training iteration based on privacy leakage risks.
However, these noise injection-based defenses struggle to balance good defense performance and model utility~\cite{xue2024revisiting, geiping2020inverting}. 
CENSOR~\cite{zhang2025censor} samples gradients from a subspace orthogonal to the original gradients while using cold Bayesian posteriors aiming to improve model utility.
\textit{Training perturbation} perturbs local training processes. 
The work~\cite{scheliga2023dropout} adds dropout layers in local models during training aiming to mitigate GIAs.
PRECODE~\cite{scheliga2022precode} adds a variational bottleneck prior to the output layer of the local model to counteract GIAs while maintaining classification performance.
The learning-rate-perturbation (LRP)~\cite{wan2023enhancing} randomly perturbs each client's learning rate to prevent accurate data reconstruction while preserving model accuracy. 
However, LRP only modifies the gradient scale without affecting the direction, making it vulnerable to strong attacks like IG that employ cosine similarity loss. 

\noindent \textbf{Pruning-based} defenses selectively remove gradient components. Prune~\cite{zhu2019deep} sets gradients with small magnitudes to zero. 
Soteria~\cite{sun2021soteria} identifies that the data representations, i.e., the outputs of the layers after the feature extractor, inferred from the gradients reveal significant information about the input data. It then prunes the selected gradients to perturb the data representations.
{ The work~\cite{zhang2023preserving} proposes pruning large gradients to defend against GIAs.} Dual Gradient Pruning (DGP)~\cite{xue2024revisiting} prunes both large and small gradients to conceal label information while incorporating an error feedback mechanism~\cite{karimireddy2019error} that adds back the pruned gradients in the next training step to mitigate information loss caused by pruning.
{
\noindent \textbf{Compression-based} defenses mitigate information leakage by compressing gradients.
$p$FGD~\cite{palihawadana2023mitigating} combines Discrete Cosine Transform and gradient pruning to suppress sensitive frequency components. 
Mixed Quantitation (MQ)~\cite{ovi2023mixed} assigns varying quantization precisions across model layers.
Existing defenses directly apply gradient compression techniques without tailoring them to defend against GIAs or considering realistic non-IID scenarios. In comparison, our truncated SVD-based defense adapts to client vulnerability, improves accuracy and defense performance via weighted approximation, and enhances aggregation effectiveness under class imbalance.
}

\subsection{Singular Value Decomposition}
\label{svd}
Low-rank approximation is the process of approximating a matrix $\vec{W}$ by a matrix ${\hat{\vec{W}}}$ of lower rank. Formally, the objective is to minimize the approximation error $||\vec{W}-\hat{\vec{W}}||$ subject to $\text{rank}(\vec{\hat{W}})\leq k$, where $k$ is the desired reduced rank.
SVD can solve this problem effectively~\cite{hsu2022language}.
For a matrix $\vec{W} \in \mathbb{R}^{p \times q}$ with $p\geq q$, SVD decomposes it as $\vec{W} = \vec{U} \vec{\Sigma} \vec{V}^{\top}$, where $\vec{U} \in \mathbb{R}^{p \times r}$ is an orthogonal matrix of left singular vectors, $\vec{\Sigma} = \text{diag}(\sigma_1,\cdots,\sigma_r) \in \mathbb{R}^{r \times r}$ contains singular values in descending order $(\sigma_1 \ge \cdots \ge \sigma_r)$, $r \leq \min \{p, q\}$ is the rank of $\vec{W}$, and $\vec{V}^{\top} \in \mathbb{R}^{r \times q}$ is the transpose of an orthogonal matrix of right singular vectors. 
Truncated SVD approximates $\vec{W}$ by retaining only the $k$ largest singular values and their corresponding singular vectors as $\hat{\vec{W}}=\vec{U}' \vec{\Sigma}' \vec{V}'^{\top}$, where $\vec{U}' \in \mathbb{R}^{p \times k}$, $\vec{\Sigma}' \in \mathbb{R}^{k \times k}$, and $\vec{V}'^{\top} \in \mathbb{R}^{k \times q}$.
The number of retained singular values is determined by the energy threshold $\mathcal{T}$, where energy refers to the sum of the squared singular values representing the amount of information captured by the singular values~\cite{leskovec2020mining}. Specifically, $k$ is chosen to satisfy $\min\limits_{k}\frac{\sum_{i=1}^k \sigma_i^2}{\sum_{i=1}^r \sigma_i^2} > \mathcal{T}$. If $pk + k + kq < pq$, the truncated matrices contain fewer parameters than the original matrix $\vec{W}$, thus reducing communication cost. 
SVD is commonly used in principal component analysis (PCA) for dimensionality reduction~\cite{shlens2014tutorial}, latent semantic analysis (LSA) for feature extraction in natural language processing~\cite{peter2009evaluation},
and addressing the challenge of sharing large-scale model updates in FL~\cite{wu2022communication, kwon2022efficient, shin2024effective}. 
Unlike prior works, we explore truncated SVD to defend against adaptive GIAs.

\section{THREAT MODEL}
\label{sec:threat-model}

\noindent \textbf{FL Setting.}
We consider a standard FL setting with a central server and $M$ distributed clients. Each client $m$ has local training data $D_m=\{(\vec{x}_{m,n},y_{m,n})\}^{N_m}_{n=1}$ with $N_m$ samples. In communication round $t$, the server first sends the global model parameters $\vec{\Theta}_g^{t-1}$ to a batch of $B^t$ selected clients and sets their local model parameters to $\vec{\Theta}_m^{t-1} = \vec{\Theta}_g^{t-1}$. Then, the clients perform local training on $D_m$ and update their respective local models from $\vec{\Theta}_m^{t-1}$ to $\vec{\Theta}_m^{t}$. 
In our setting, each client $m$ transmits its local model gradients $\vec{\nabla\Theta}_m$ to the server, which has the same effect as sending the updated local model parameters since $\vec{\nabla\Theta}_m = \vec{\Theta}_g^{t-1} - \vec{\Theta}_m^t$. Finally, the server updates the global model as $\vec{\Theta}_g^t = \vec{\Theta}_g^{t-1} -\sum_{b=1}^{B^t} p_b \vec{\nabla\Theta}_b$, where $p_b$ is the normalized aggregation weight of the $b$-th selected client.
The objective is to collaboratively train a global model $\vec{\Theta}_g$ by aggregating client updates. We follow the FedAvg algorithm~\cite{mcmahan2017communication} and formulate the objective as $\arg\min\limits_{\vec{\Theta}_g} \sum\limits_{m=1}^M \sum\limits_{n=1}^{N_m} \mathcal{L}(F_{\vec{\Theta}_g}(\vec{x}_{m,n}),y_{m,n})$, where $F_{\vec{\Theta}_g}(\cdot)$ is the neural network with parameters $\vec{\Theta}_g$ and $\mathcal{L}(\cdot, \cdot)$ is the loss function used to train it. 

\noindent \textbf{Adversary Model and Capabilities.}
We consider an \textit{honest-but-curious} central server as the adversary. Such an adversary is common in FL research~\cite{zhu2019deep,geiping2020inverting,sun2021soteria,wang2023more}. This adversary follows the FL protocol properly but attempts to reconstruct clients' private training data from their uploaded updates. 
The adversary has access to the global model parameters $\vec{\Theta}_g$ and local model gradients $\vec{\nabla \Theta}_m$
but cannot modify the training process or tamper with model parameters. Furthermore, we assume an \textit{adaptive adversary} who knows the deployed defense mechanisms and can design targeted attacks against these defenses. 
We assume that the adversary has enough computational resources to perform attacks. 


\noindent \textbf{Adversary Goals.} The adversary aims to reconstruct private data from local model gradients via GIAs.
Given client $m$'s gradients $\vec{\nabla \Theta}_m$,
the adversary first initializes pairs of dummy input data $\vec{x}_m^{\prime}$ and label $y_m^{\prime}$, which are the optimizable parameters for data recovery. 
After forward and backward propagation on the global model, the dummy gradients $\vec{\nabla \Theta}^{\prime}_m = \nabla \mathcal{L}(F_{\Theta_g}(\vec{x}_m^{\prime}),y_m^{\prime})$ can be generated. The reconstruction of private data $\vec{x}_m$ can be viewed as an iterative optimization process with the objective of minimizing the distance between the dummy gradients and the ground-truth gradients of the victim client, which can be formulated as $\underset{\vec{x}_m^{\prime}, y_m^{\prime}}{\arg\min}\, \text{Dist}(\vec{\nabla \Theta}^{\prime}_m, \vec{\nabla \Theta}_m) + \mathcal{R}(\vec{x}_m^{\prime})$, where $\mathcal{R}(\cdot)$ is the regularization term. 

\section{MOTIVATION STUDY}
\label{sec:motivation-study}

{ This section presents a preliminary study on the MNIST dataset~\cite{deng2012mnist}, demonstrating that existing GIA defenses can be circumvented by practical adaptive adversaries. We summarize key insights from our study to motivate our defense design.}

\noindent \textbf{Experimental Setup}.
We consider five representative defenses, including two perturbation-based defenses, \textit{CENSOR}~\cite{zhang2025censor} and \textit{PRECODE}~\cite{scheliga2022precode},  
and three pruning-based defenses, \textit{Prune}~\cite{zhu2019deep}, \textit{Soteria}~\cite{sun2021soteria}, and \textit{DGP}~\cite{xue2024revisiting}. 
In this section, we omit perturbation-based defenses that rely on noise injection, including DP~\cite{zhu2019deep} and Outpost~\cite{wang2023more}. This is because Expectation over Transformation (EoT), a practical adaptive attack operation commonly considered for mitigating random effects induced by the defenders \cite{zhang2025censor,athalye2018synthesizing,tramer2020adaptive,athalye2018obfuscated}, is ineffective against these noise injection-based defenses, as theoretically analyzed in Appendix~\ref{subsec:analysis-gradient-perturb}. 
However, we show (in \sect\ref{subsec:defense-performance}) that under a more powerful, thus less practical, adaptive adversary, our defense still outperforms the existing defenses, which demonstrates the superior performance of our proposed solution even in stressful situations.
We configure the defenses following the recommended settings in their respective publications to reproduce their best performance.
For CENSOR, we activate the defense for the first five epochs, following ~\cite{zhang2025censor}.
For PRECODE, we adopt the same parameter settings as in~\cite{scheliga2022precode}. For Prune, we set the pruning rate to 90\%. For Soteria, we prune 80\% of the gradients in the fully connected layers. For DGP, we prune the smallest 75\% as well as the largest 5\% of the gradients. 

We evaluate the defenses against both non-adaptive and adaptive attacks. For the non-adaptive scenarios, we use the standard IG attack. 
For the adaptive scenarios, we apply defense-specific operations on the original IG attack. 
For CENSOR, the adversary is assumed to know that the defense is active for the initial epochs. We then perform the attack in the subsequent undefended epochs and report the best attack results. Note that if the defense keeps active for all the epochs in CENSOR, the model's utility will degrade significantly, as shown in \sect\ref{acc}.
For PRECODE, we initialize a dummy random vector and optimize it together with the dummy inputs during attack optimization. 
For the pruning-based defenses (i.e., Prune, Soteria, and DGP), we assume that the adversary knows the defense and details of the pruning operation by detecting the zero values in the ground-truth gradients. We then apply identical pruning operations to both the ground-truth and dummy gradients during attack optimization. 
Since these defense mechanisms and their parameters (e.g., the number of initial epochs where defense is deployed) are static information, the adversary can obtain them in an advanced persistent threat (APT) scenario~\cite{alshamrani2019survey}, where the adversary may use, for instance, social engineering against the FL clients to exfiltrate the needed knowledge. 
{ Note that we omit GAN-based attacks in this discussion because they yield results similar to optimization-based attacks (validated in \sect\ref{subsec:defense-performance}) due to the two approaches' shared optimization objective, as discussed in \sect\ref{sec:threat-model}. Consequently, the adaptive operations in this section are compatible with both attack types.}

\begin{table}[t]
    \caption{Defense Performance of Different Methods Under Non-adaptive and Adaptive GIAs. Higher PSNR and Lower LPIPS Values Mean Stronger Attack Performance.
    }
    \centering
    \begin{tabular}{|c|c|c|c|c|}
    \hline
    \multirow{2}{*}{\diagbox{Defense}{Metric}} & \multicolumn{2}{c|}{Non-adaptive} & \multicolumn{2}{c|}{Adaptive} \\
      
    \cline{2-5}
     & PSNR & LPIPS & PSNR & LPIPS \\
     \hline
     CENSOR~\cite{zhang2025censor} & 8.1940 & 0.6958 & 16.4071 & 0.2881 \\
  
    \hline
    PRECODE~\cite{scheliga2022precode} & 3.5659 & 0.7668 & 57.4165 & 0.0001 \\
    \hline
    Prune~\cite{zhu2019deep} &12.9257 & 0.4993 & 36.1273 & 0.0221 \\
    \hline
    Soteria~\cite{sun2021soteria} & 10.8145 & 0.6481 & 38.7447 & 0.0161 \\
    \hline
    DGP~\cite{xue2024revisiting} & 9.7334 & 0.6187 & 35.6383 & 0.0254 \\
    \hline
    \end{tabular}
    
    \label{tab:adaptive-PSNR-LPIPS}
\end{table}

\begin{figure}[t!]
\centering
\includegraphics[width=\columnwidth]{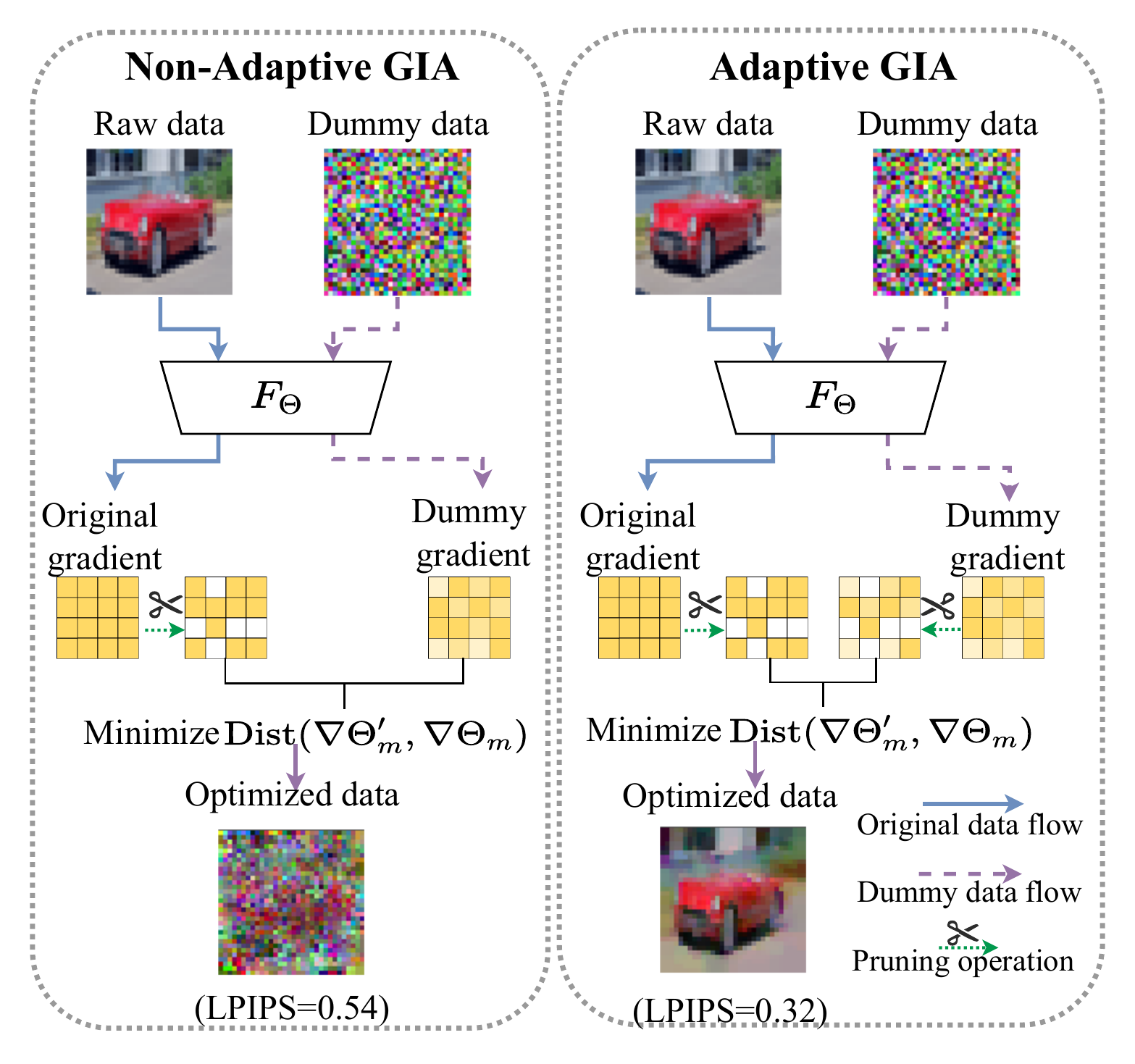}
\centering
\caption{Comparison of non-adaptive and adaptive GIAs against the Prune defense~\cite{zhu2019deep}.
}
\label{adaptive_attack}
\end{figure}

\noindent \textbf{Experimental Results.}
Table~\ref{tab:adaptive-PSNR-LPIPS} summarizes the peak signal-to-noise ratio (PSNR)~\cite{gonzalez2020digital} and the learned perceptual image patch similarity (LPIPS)~\cite{zhang2018unreasonable} between the ground-truth and reconstructed images under the non-adaptive and adaptive GIAs. Higher PSNR values mean better reconstruction quality and lower LPIPS values mean smaller perceptual differences between the original and reconstructed images, both indicating stronger attack performance. 
Although LRP is not effective against the strong IG attack, we still show that its defense performance drops under the adaptive DLG attack \cite{zhu2019deep} in Appendix~\ref{lrp}.
These results demonstrate that the effectiveness of the subject defenses drops significantly under adaptive attacks.
Fig.~\ref{adaptive_attack} exemplifies how an adaptive attack works against the Prune defense. From the figure, while the non-adaptive attack directly matches gradients without considering the defense, the adaptive attack applies the same pruning operation to both the ground-truth and dummy gradients during optimization, thereby achieving better reconstruction performance.

\noindent \textbf{Key Insights}.
{ Our experimental results demonstrate key vulnerabilities of the existing defenses under practical adaptive adversaries.
First, pruning-based defenses can be adaptively attacked by identifying zeroed gradient components and exploiting the unaffected ones for reconstruction. Second, defenses relying on random variables (e.g., PRECODE and LRP) can be bypassed through variable recovery. These empirical findings suggest that affecting all the gradient components and doing so irreversibly are desirable properties for improved robustness against adaptive GIAs.
Third, by applying random orthogonal projection to gradients in the initial epochs only, CENSOR is vulnerable to attacks during the later undefended epochs, where the adversary can still extract sufficient information to reconstruct much of the input data, as evidenced by our experiments. If the random projection were to be kept active to maintain privacy, the totality of projected gradients would retain so little information that it significantly degrades the model's utility.
Therefore, it is essential for practical defenses to achieve both privacy protection and good model utility.
These insights motivate our proposed solution based on truncated SVD.} 
On the one hand, truncated SVD irreversibly affects all the gradient components.
On the other hand, it prudently truncates the gradients while preserving critical information for model utility at low computational and communication overheads. However, applying truncated SVD as a robust GIA defense presents practical challenges, which we will address in the next section.


\section{System Design}
\label{sec:system-design}

\subsection{Challenges and Design Goals}
\label{subsec:challenges}

In real-world FL deployments, data distributions among clients are often not independent and identically distributed (non-IID). One of the most common non-IID scenarios is the imbalanced distribution of classes~\cite{deng2022tailorfl,hsieh2020non,xu2024overcoming}, where clients possess varying proportions of data samples across different classes. For example, different hospitals may observe different frequencies of disease types based on their specialties and patient demographics. While prior efforts~\cite{deng2022tailorfl,hsieh2020non,xu2024overcoming,lu2024federated} have focused on improving model utility under non-IID data in FL, the impact of such data heterogeneity on gradient inversion attacks and defenses remains unexplored. 

\begin{figure}[t!]
\centering
\includegraphics[width=.9\columnwidth]{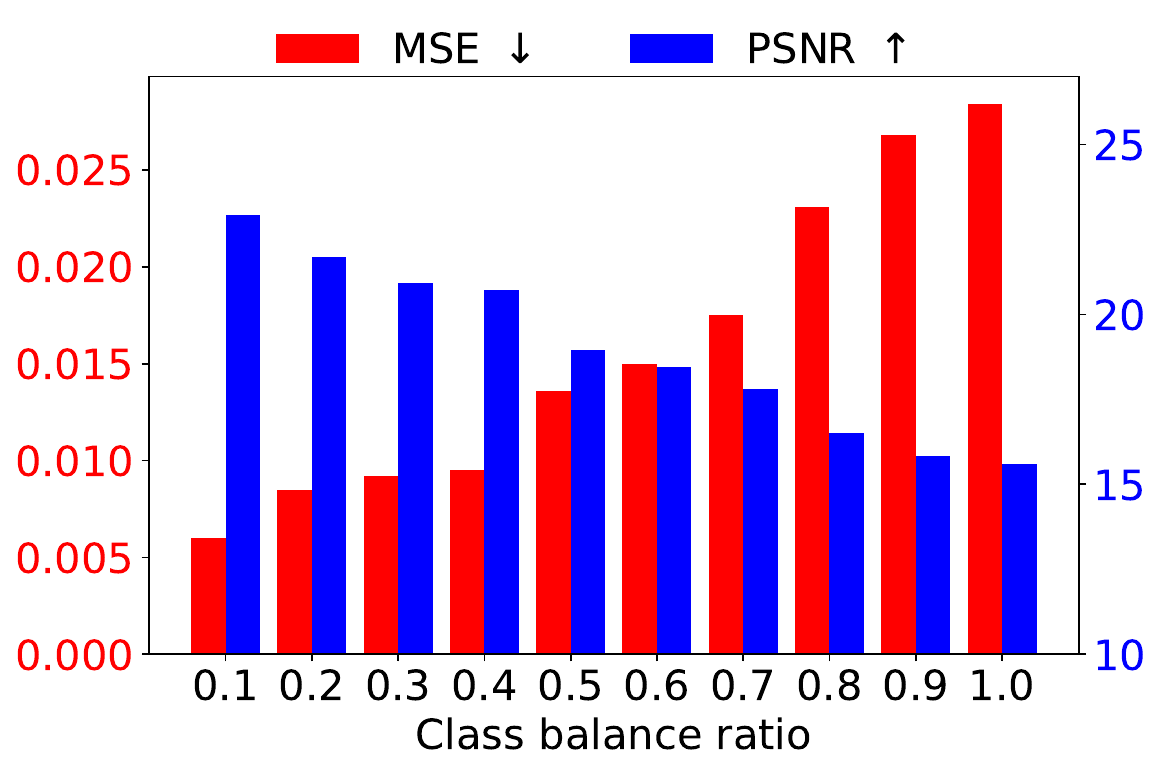}
\centering
\caption{Impact of class imbalance on attack effectiveness. Lower MSE and higher PSNR indicate stronger attack performance.
}
\label{non_iid}
\end{figure}

In our investigation, we simulate the non-IID scenario of class imbalance using the MNIST dataset and adopt ResNet-18~\cite{he2016deep} as our target model. We first randomly shuffle the order of the data classes and retain $N_{cls\_i} \times \rho^{cls\_i}$ samples for each class, where $cls\_i$, ($cls\_i = 0, \cdots 9$) is the shuffled class index, $N_{cls\_i}$ is the original number of samples in class $cls\_i$, and $\rho$ is our defined \textit{class balance ratio}. By varying $\rho$ from 0 to 1 with a step size of 0.1, we simulate different degrees of class imbalance, where a larger $\rho$ indicates a more balanced class distribution. For each $\rho$, we generate 128 batches of input samples with a batch size of 10. Then, we launch the IG attack.
As shown in Fig.~\ref{non_iid}, as $\rho$ increases, the mean squared error (MSE) between the original and reconstructed images increases and the PSNR value decreases. This suggests that clients subject to higher degrees of class imbalance are more vulnerable to attacks.
This is potentially because, when trained on class-imbalanced data, the model's gradients primarily reflect patterns from the dominant classes. Therefore, it becomes easier for attackers to reconstruct private data from the less diverse gradients.

Based on the above observations, it is inadvisable to treat clients with varying degrees of class imbalance uniformly when counteracting GIAs through SVD truncation. Since more imbalanced clients are more vulnerable to attacks, they require stronger protection through lower energy thresholds that consequently reduce available information in the truncated gradients for data reconstruction, as illustrated in Appendix~\ref{threshold_mse}.
This strategy raises three practical challenges during SVD truncation: \textbf{Challenge C1:} How to effectively quantify the degree of class imbalance and thereby adaptively adjust the energy threshold $\mathcal{T}$ under heterogeneous clients; \textbf{Challenge C2:} How to preserve information critical for model training while suppressing sensitive information leakage; and \textbf{Challenge C3:} How to effectively aggregate the SVD truncated client updates and improve global model utility under class imbalance? Our design aims to address these three challenges.

\subsection{Overview of \textit{SVDefense}}
\begin{figure*}[t!]
\centering
\includegraphics[width=\textwidth]{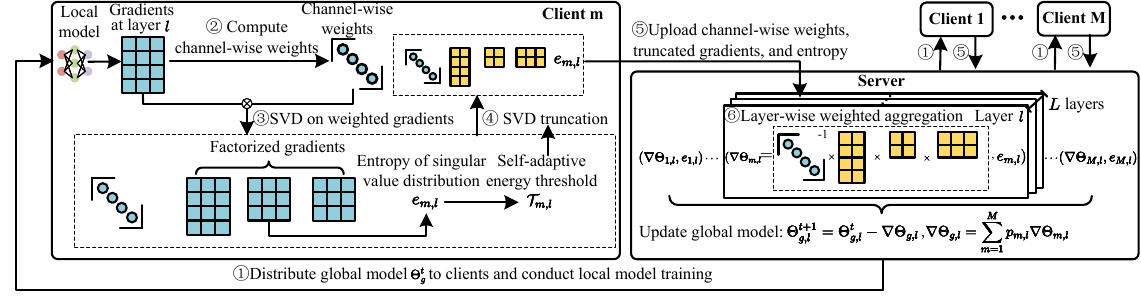}
\centering
\caption{
Overview of the proposed \textit{SVDefense}. 
{ 
On the server side, $\nabla \vec{\Theta}_{m,l}$ denotes client $m$'s reconstructed gradients at layer $l$, $\nabla \vec{\Theta}_{g,l}$ is the global gradients at layer $l$, and $p_{m,l}$ represents the layer-wise aggregation weight. 
}
}
\label{fig:overview}
\end{figure*}

\begin{figure}[t!]
    \centering
    \begin{minipage}{0.4\columnwidth} 
        \centering
        \includegraphics[width=\linewidth]{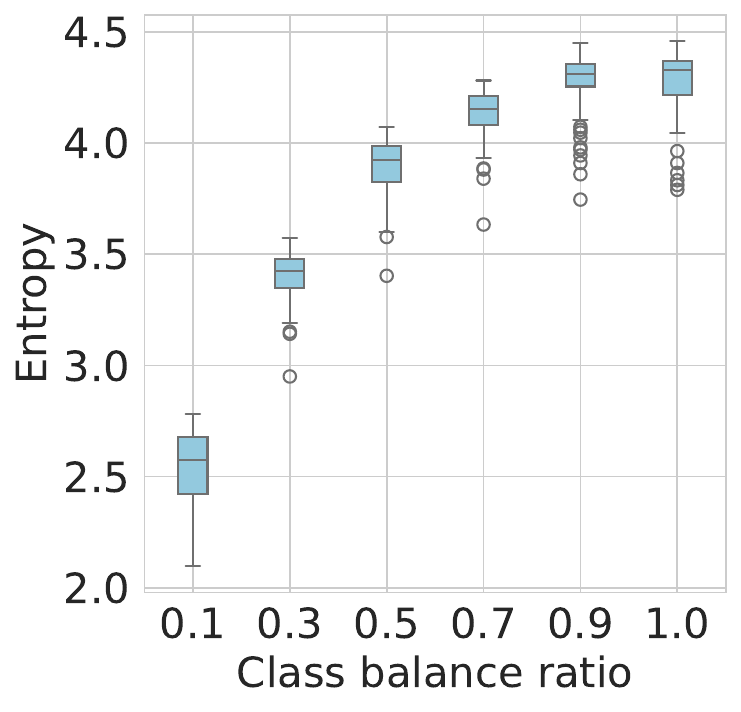} 
        \caption{Entropy of singular value distribution vs. class balance ratio.}
        \label{fig:entropy_variation_right}
    \end{minipage}
    \label{fig_sim}
    \hfill
        \begin{minipage}{0.55\columnwidth}
        \centering
        \includegraphics[width=\linewidth]{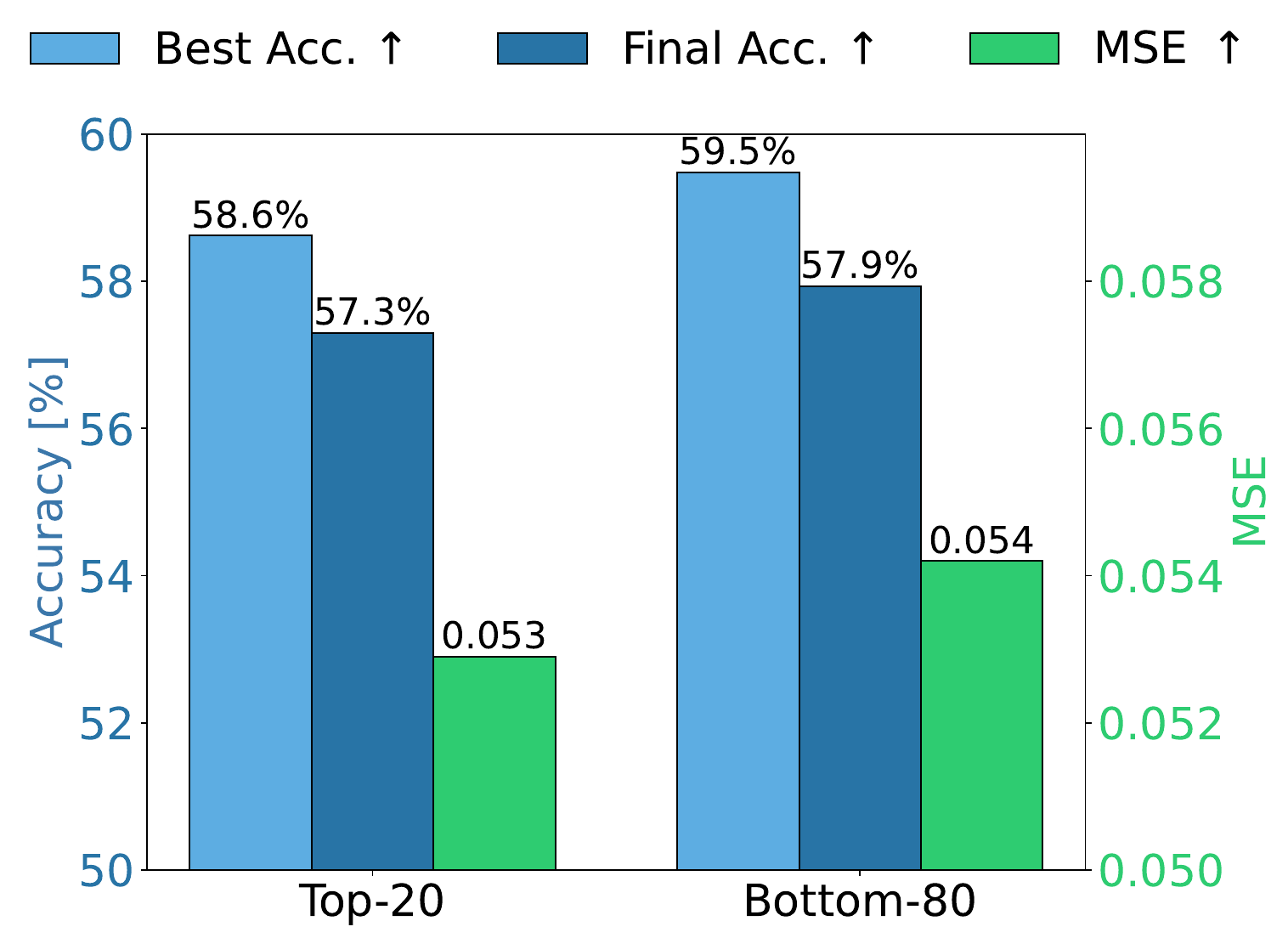} 
        \caption{Comparison of different gradient perturbation-based defense strategies under GIAs.}
        \label{toy_noise}
    \end{minipage}

\end{figure}

Fig.~\ref{fig:overview} gives an overview of the proposed \textit{SVDefense}. { We also provide a detailed description of our approach in Alg.~\ref{alg:svdefense}. The workflow is as follows.}
\textcircled{1} Each client receives the global model and trains its local model { (lines 4-6)}.
\textcircled{2} Each client computes channel-wise weights based on its gradient magnitude information { (line 17)}. 
\textcircled{3} Each client applies the channel-wise weights to perform SVD on the gradients,
measures the entropy of the squared singular value distribution, and derives the self-adaptive energy threshold { (lines 18-20)}.
\textcircled{4} The factorized gradients are then truncated according to the calculated energy threshold for each client { (line 21)}.
\textcircled{5} Each client transmits channel-wise weights, truncated gradients, and entropy value to the server { (line 9)}. 
\textcircled{6} The server reconstructs the local gradients with the truncated gradients { (line 26)} and calculates layer-wise aggregation weights based on the entropy value { (line 27)} to update the global model { (line 13)}.

\begin{algorithm}[t] 
\caption{ Federated learning with \textit{SVDefense}}
\label{alg:svdefense}
\begin{algorithmic}[1]
\State \textbf{Input:} Initial global model $\vec{\Theta}_g^0$, local datasets $\{D_m\}_{m=1}^M$, number of rounds $T$, sensitivity parameter $\beta$
\State \textbf{Output:} Final global model $\vec{\Theta}_g^{T}$
\For{each round $t = 0, 1, \ldots, T-1$}
    \For{each selected client $m$ in parallel}
        \State Initialize local model: $\vec{\Theta}_m \gets \vec{\Theta}_g^t$
        \State Train local model on $D_m$ to get gradients $\nabla \vec{\Theta}_m$
        \For{each layer $l$ of client model}
        \State $\vec{P}_{m,l} \gets \textsc{Defend\_Grad}(\nabla \vec{\Theta}_{m,l}, \beta)$ 
        \State Send $\vec{P}_{m,l}$ to server
        \EndFor
    \EndFor
    \For{each layer $l$ of global model in server}
    \State $\vec{\Theta}_{g,l}^{t+1} \gets \textsc{Aggregate}(\vec{\Theta}_{g,l}^t, \{\vec{P}_{m,l}\}_{m=1}^M)$
    \EndFor
\EndFor

\Function{{Defend\_Grad}}{$\nabla \vec{\Theta}_{m,l}, \beta$}
    \State Compute channel-wise weight matrix $\vec{I}_{l}$ from $\nabla \vec{\Theta}_{m,l}$
    \State $\{\vec{U}_{l}, \vec{\Sigma}_{l}, {\vec{V}_{l}}^\top\} \gets \text{SVD}(\vec{I}_{l} \nabla \vec{\Theta}_{m,l})$
    \State Compute entropy $e_{m,l}$ from $\vec{\Sigma}_{l}$
    \State Compute threshold $\mathcal{T}_{m,l} \gets 1 - \text{exp}( - \beta e_{m,l})$
    \State Truncate singular values to get $\{\vec{U}^{*}_{l}, \vec{\Sigma}^{*}_{l}, {\vec{V}_{l}^{*}}^\top\}$ 

    \State \Return $\vec{P}_{m,l} \gets\{\vec{I}_l, \vec{U}_l^{*}, \vec{\Sigma}_l^{*}, {\vec{V}_l^{*}}^\top, e_{m,l}\}$
\EndFunction

\Function{Aggregate}{$\vec{\Theta}_{g,l}^t,\{\vec{P}_{m,l}\}_{m=1}^M$}

    \For{each client $m$}
        \State $\nabla \vec{\Theta}_{m,l} \gets {(\vec{I}_{m,l})}^{-1} \vec{U}_{m,l}^{*} \vec{\Sigma}_{m,l}^{*} {\vec{V}_{m,l}^{*}}^\top$
        \State Compute layer-wise aggregation weight $p_{m,l}$ based on $e_{m,l}$
    \EndFor
    \State $\nabla \vec{\Theta}_{g,l} \gets \sum^{M}_{m=1} p_{m,l} \vec{\nabla \Theta}_{m,l}$
    \State \Return $ \vec{\Theta}_{g,l}^t - \nabla \vec{\Theta}_{g,l}$
\EndFunction

\end{algorithmic}
\end{algorithm}

\subsection{Self-Adaptive Energy Threshold}
\label{self_adaptive}

Our analysis in \sect\ref{subsec:challenges} indicates that the clients with higher degrees of class imbalance require stronger protection against GIAs by lowering the energy threshold. To address \textbf{Challenge C1} and quantify the degree of class imbalance for adapting each client's energy threshold, we follow the experimental protocol in \sect\ref{subsec:challenges} and apply SVD to decompose the gradients obtained at the end of each training epoch for each setting of $\rho$. Fig.~\ref{fig:entropy_variation_right} shows the entropy of the squared singular value distribution for a linear layer of the target model versus the class balance ratio $\rho$. We can see that the entropy value increases with the class balance ratio, indicating that the singular value distribution can serve as an effective indicator of the degree of class imbalance. 
This observation motivates us to adapt the energy threshold $\mathcal{T}_{m,l}$ for layer $l$ of client $m$'s local model based on the entropy $e_{m,l}$ of the normalized squared singular values.
Specifically, we set

\begin{equation}
\label{equ:adaptive_threshold}
\begin{aligned}
& \mathcal{T}_{m,l}  = 1 - \text{exp}( - \beta e_{m,l}),
\end{aligned}
\end{equation}
where $e_{m,l} = -\sum_{i=1}^{r_{m,l}} \tilde{\sigma}_{m,l,i} \mathrm{log}(\tilde{\sigma}_{m,l,i})$ with $\tilde{\sigma}_{m,l,i}=\frac{({{\sigma}_{m,l,i}})^2}{\sum_{j=1}^{r_{m,l}} ({{\sigma}_{m,l,j}})^2}$ as the $i$-th normalized squared singular value, $r_{m,l}$ is the rank of the gradients of layer $l$ in client $m$'s model, and the sensitivity parameter $\beta$ controls the sensitivity of privacy protection. A larger $\beta$ means a faster decrease in threshold values as the entropy decreases. This exponential function ensures that clients with more imbalanced class distributions (indicated by lower entropy values) receive lower energy thresholds, resulting in stronger privacy protection. 



\subsection{Channel-Wise Weighted Approximation} 


Lowering the energy threshold $\mathcal{T}$ in SVD truncation not only suppresses sensitive information that can be exploited for data reconstruction, but it also reduces useful information for training the model. 
Traditional SVD treats all the elements of the matrix uniformly, which does not align with our objective of addressing
\textbf{Challenge C2} to preserve more information critical for model training while suppressing sensitive information leakage. 
Recent studies~\cite{zhu2019deep,sun2021soteria,xue2024revisiting} suggest that larger gradients, which capture the primary direction of model updates, contain more critical information for classification, while smaller gradients often carry redundant information. 
Additionally, as proven in \cite{xue2024revisiting}, the effectiveness of GIAs measured by the data reconstruction error is bounded by the overall gradient error, regardless of whether it originates from large or small gradients.
Inspired by these insights, our design preserves larger gradients while applying stronger perturbations to smaller gradients during SVD truncation, 
aiming to improve both defense performance and model utility.

We conduct a toy experiment on CIFAR-10 using the IG attack to illustrate the effectiveness of the aforementioned strategy. We consider two perturbation-based defenses: 1) \textit{top-20} that applies Laplace noise (scale = 0.03) to the top 20\% largest gradients and 2) \textit{bottom-80} that applies Laplace noise (scale = 0.03) to the bottom 80\% smallest gradients. As shown in Fig.~\ref{toy_noise}, the bottom-80 defense strategy achieves higher best and final classification accuracies and higher MSE between the ground truth and reconstructed inputs, 
indicating a better classification and defense performance at the same time.



To this end, we propose a weighted truncated SVD approach that incorporates the gradient magnitude information.
Intuitively, the weighted optimization objective is formulated as:
$
    \arg\min\limits_{\hat{\vec{W}}_l} \sum_{c=1}^C\sum_{i=1}^{N_{l,c}} \omega_{l,c, i} ( \vec{W}_{l,c,i}-\hat{\vec{W}}_{l,c,i})^2,
$
where $\vec{W}_l$ and $\hat{\vec{W}}_l$ denote respectively the ground-truth and approximated gradients of the $l$-th layer in the client model, $\omega_{l,c, i}$ is the weight for the $i$-th element in the $c$-th output channel of the gradients, $C$ is the number of total output channels, and $N_{l,c}$ is the number of elements in the $c$-th channel of $\vec{W}_l$. 
For simplicity, we use squared gradients as weights, { where $\omega_{l,c,i} = (\vec{W}_{l,c,i})^2$}. 
In { \sect\ref{subsec:exp-channel-wise}}, we consider an alternative method of using absolute gradient values as weights and demonstrate that squared gradients as weights achieves better accuracy and defense performance.
However, such element-wise weighted low-rank approximation is a nonlinear optimization problem, which does not have a closed-form solution~\cite{srebro2003weighted}. 
To make the problem tractable, we employ a diagonal weight matrix.
Specifically, we sum the weights along the output channel axis and
define the channel-wise weight matrix as
$\vec{I}_l = \text{diag}(
\sqrt{ \omega_{l,1}},\ldots, \sqrt{ \omega_{l,C}})$, where $\omega_{l,c} = \sum_{i=1}^{N_{l,c}} \omega_{l,c, i}$. The optimization problem then transforms to:
\begin{equation}
    \arg\min\limits_{\hat{\vec{W}}_l} \| \vec{I}_l\vec{W}_l -\vec{I}_l\hat{\vec{W}}_l\|_F,
\label{obj}
\end{equation}
where $\|\cdot\|_F$ is the Frobenius norm. The optimization tends to reduce the error in high-weight regions due to $\vec{I}_l$. 

The optimization process is as follows. For a rank $k$ derived from the energy threshold $\mathcal{T}$, we obtain the optimal $\hat{\vec{W}}^{*}_l$ by applying truncated SVD to $\vec{I}_l\vec{W}_l$, yielding $\vec{U}^{*}_l$, $\vec{\Sigma}^{*}_l$, and ${\vec{V}^{*}_l}^{\top}$ that satisfy Eq. \ref{obj}. The optimal solution is denoted by $\hat{\vec{W}^{*}_l} = {(\vec{I}_{l})}^{-1} \vec{U}^{*}_l \vec{\Sigma}^{*}_l {\vec{V}^{*}_l}^{\top}$, which maintains the same rank $k$ since ${(\vec{I}_{l})}^{-1}$ is a diagonal matrix.
This weighted optimization strategy effectively improves both the model accuracy and the privacy protection by selectively preserving larger gradients while applying stronger perturbations to smaller gradients. 

We theoretically demonstrate that our Channel-Wise Weighted Approximation enhances the defense performance of truncated SVD. 
{
A \textit{passive attacker} is defined as an adversary who attempts to reconstruct private input data while honestly adhering to the FL protocol~\cite{xue2024revisiting}. As discussed in \sect\ref{sec:threat-model}, the honest-but-curious adversary we consider belongs to the class of passive attackers.
}
Specifically, we have the following definition and theorem:

\noindent \textbf{Definition 1}. \textit{A passive attack $\mathcal{A}$ is an $(\varepsilon,\delta)$-passive attack, if it satisfies}:
\begin{equation}
    \mathbb{P}(\mathbb{E}(\mathcal{D}_\mathcal{A}(\vec{\nabla \Theta},\vec{\nabla \Theta^{*}})) \leq \varepsilon) \geq 1 - \delta.
\end{equation} 
\textit{where $\mathbb{P}$ represents the probability, $\mathbb{E} $ represents the expectation, and $\mathcal{D}_\mathcal{A}$ is the distance estimated under $\mathcal{A}$}.

\noindent \textbf{Theorem 1}. \textit{For any $(\varepsilon,\delta)$-passive attack $\mathcal{A}$, under the presence of truncated SVD, it will degenerate to $(\varepsilon+\sqrt{\gamma_1} \|\vec{\nabla \Theta} \|_F,\delta)$, where $\gamma_1 = 1 -\mathcal{T}$. Under the presence of truncated SVD with Channel-Wise Weighted Approximation, it will degenerate to $(\varepsilon+\sqrt{\gamma_2} \|\vec{\nabla \Theta} \|_F,\delta)$, where $\gamma_2 = (\frac{\sigma_{max}(\vec{I})}{\sigma_{min}(\vec{I})})^2 (1-\mathcal{T})$, $\sigma_{max}(\cdot)$ and $\sigma_{min}(\cdot)$ mean the maximum and minimum singular value of the input matrix.}

Since $\frac{\sigma_{max}(\vec{I})}{\sigma_{min}(\vec{I})} \geq 1$, we have $\gamma_2 \geq \gamma_1$. This indicates that truncated SVD with Channel-Wise Weighted Approximation provides stronger protection than the original truncated SVD.
The detailed proof can be found in Appendix~\ref{app:proof-defense-performance}.
By our proof in Appendix~\ref{app:proof-defense-performance} and experimental results in \sect\ref{eval}, Channel-Wise Weighted Approximation improves both the defense performance and model accuracy.


\subsection{Layer-Wise Weighted Aggregation}
\label{subsec:layer_wise}
{ Under non-IID local data distributions, the aggregated local optima deviates from the global optimum, leading to degraded global accuracy~\cite{wan2023enhancing,xu2024overcoming,karimireddy2020scaffold,chen2023elastic}.}
To address this, weighted aggregation strategies have been proposed~\cite{chen2023elastic,ahmad2023robust}, where local clients contribute differently to the global model updates based on their heterogeneous data distributions. However, no existing work investigates appropriate weights for aggregating SVD truncated weights and addresses \textbf{Challenge C3}.

{ This work assumes the global data distribution to be class-balanced, which aligns with standard FL benchmarks~\cite{cohen2017emnist,krizhevsky73cifar,anguita2013public,warden2018speech}. Under this assumption, clients with more balanced local data distributions will have local optima closer to the global optimum, which should be assigned higher weights during aggregation~\cite{xu2024overcoming}.}
Our analysis in \sect\ref{self_adaptive} reveals that the entropy of the squared singular value distribution can serve as an indicator of the degree of class imbalance. Therefore, we assign layer-wise aggregation weights to different clients as:

\begin{equation}
\label{server_weight}
    p_{m,l} = \frac{e_{m,l} \times N_m}{\sum\limits_{i=1}^M e_{i,l} \times N_i}, 
\end{equation}
where $ p_{m,l}$ represents the aggregation weight.
The server then aggregates the received client updates to update the global model $\vec{\Theta}_{g,l}^{t+1} = \vec{\Theta}_{g,l}^{t} - \vec{\nabla \Theta}_{g,l}$, where $\vec{\Theta}_{g,l}^t$ is the weights of layer $l$ of the global model, $\vec{\nabla \Theta}_{g,l} = \sum^{M}_{m=1} p_{m,l} \vec{\nabla \Theta}_{m,l}$,
and $\vec{\nabla \Theta}_{m,l}$ denotes the layer $l$ of client $m$'s gradients.



\section{Evaluation}
\label{eval}
\subsection{Experimental Setup.}
\label{subsec:eval-setup}

\begin{table}[t!]
\caption{Number of Total and Participating Clients per Round.}
\resizebox{\linewidth}{!}{
\begin{tabular}{ccccc}
\hline \toprule
Dataset& \# total client & \makecell{\# client \\ per round} & \# class & non-IID \\
\hline
EMNIST~\cite{cohen2017emnist}&        196 & 19 &       
64 & \checkmark\\
       CIFAR-10~\cite{krizhevsky73cifar}&        100 &10&        10 &\checkmark \\
       
 HAR~\cite{anguita2013public}& 30 &2&6 & \checkmark\\
 KWS~\cite{warden2018speech} & 489 &48&36 & \checkmark\\
 \hline \toprule
 \end{tabular}
}
\label{tab:num-client}
\end{table}

\begin{figure*}[t!]
  \centering
  \subfloat[IC-CIFAR10]{
    \centering
    \includegraphics[width=0.47\columnwidth]{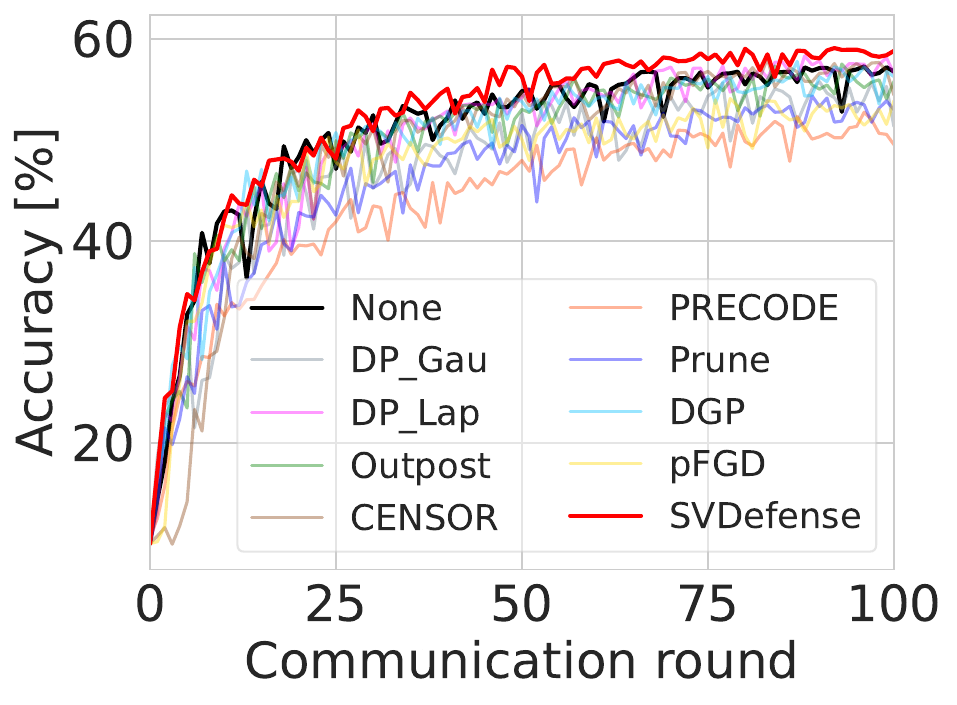}
    \label{fig:acc_cifar}
    }
  \hfill
  \centering
  \subfloat[IC-EMNIST]{
    \centering
    \includegraphics[width=0.47\columnwidth]{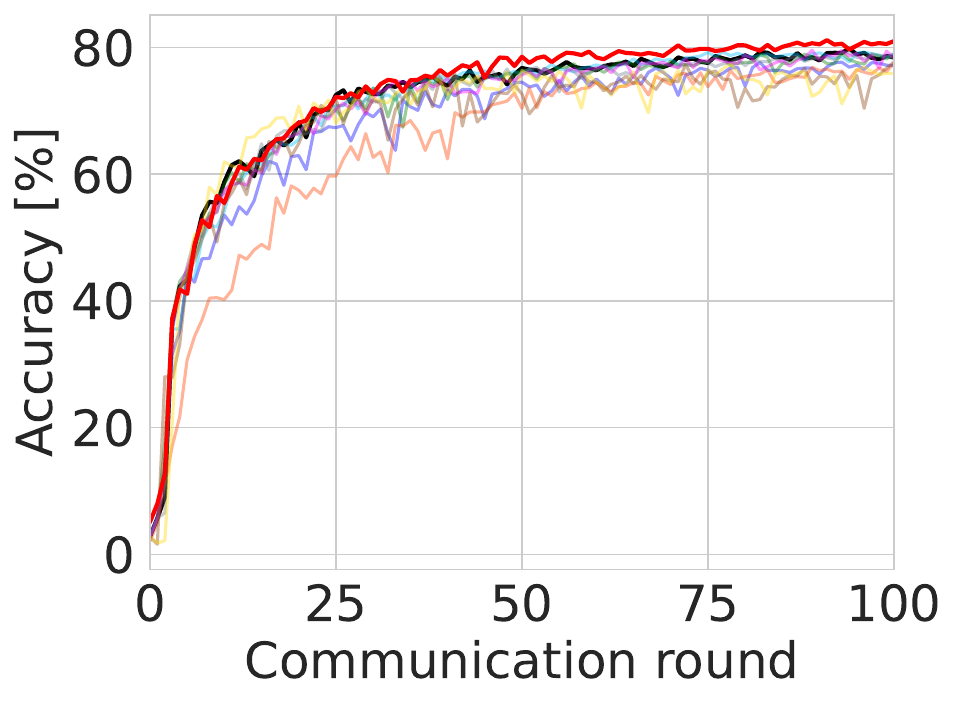} 
  }
   \hfill
  \centering
  \subfloat[HAR]{
    \centering
    \includegraphics[width=0.47\columnwidth]{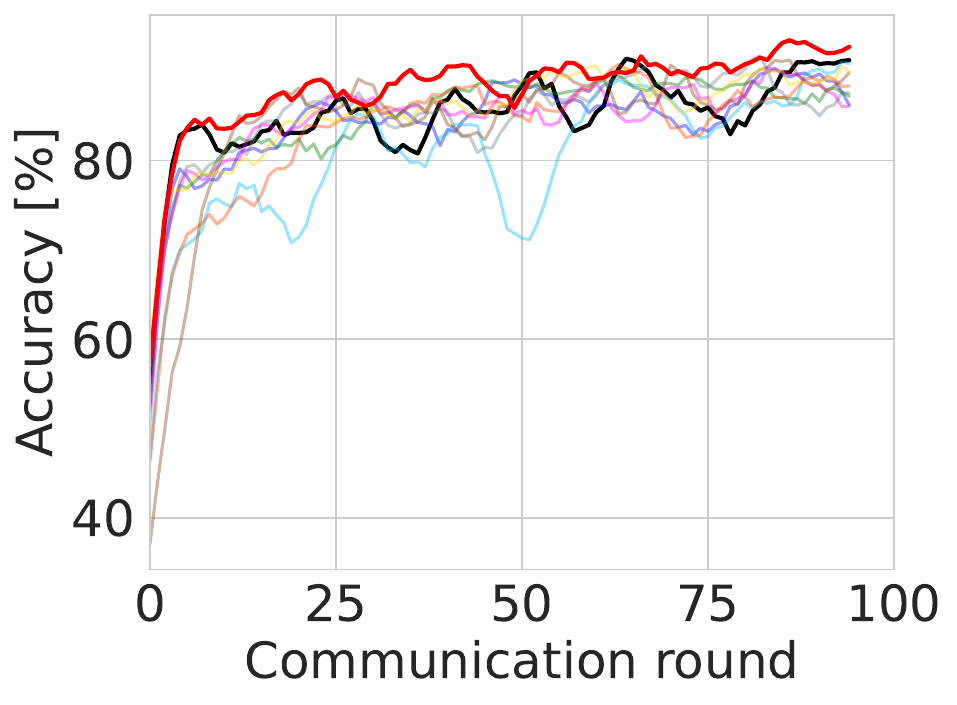} 
  }
  \hfill
  \subfloat[KWS]{
    \centering
    \includegraphics[width=0.47\columnwidth]{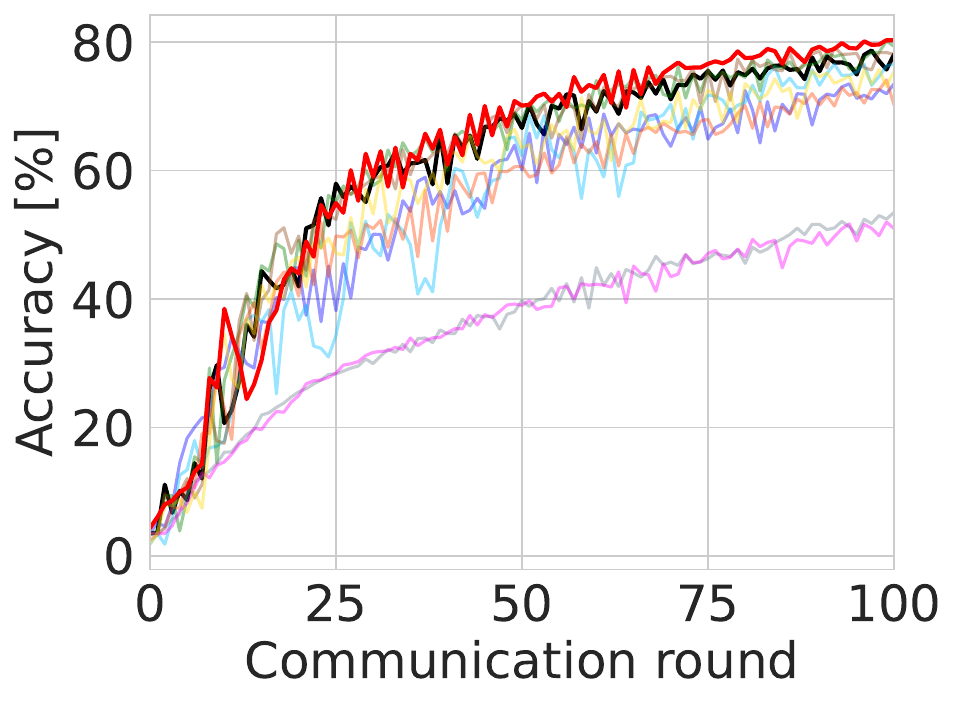} 
  }
  \caption{
  Comparison of classification accuracy across different defense methods.}
  \label{fig_acc}
\end{figure*}

\begin{table*}[t!]
\caption{Comparison of Defense Effectiveness Across Different Defense Methods.}
\label{tab:defense}
\centering
\resizebox{\textwidth}{!}{
\begin{tabular}{ccccccccccccc}
\hline
\toprule
Dataset & Metric & None & DP-Gau & DP-Lap & Outpost  &CENSOR& PRECODE & Prune & Soteria & DGP & { $p$FGD} &\textit{SVDefense} \\ \hline
\multirow{4}{*}{CIFAR-10} 
& MSE $(\uparrow)$ & 0.0056 & {0.0546}& 0.0514 & 0.0177  &0.0141& 0.0000 & 0.0136 & 0.0050 & 0.0108 & { \underline{0.0584}}&\textbf{0.0619} \\
& PSNR $(\downarrow)$ & 23.8755 & \underline{12.8280} & 13.1080 & 18.0419  &19.2682& inf & 19.3477 & 24.0950 & 21.1468 & { 12.9291}& \textbf{12.5278} \\
& SSIM $(\downarrow)$ & 0.8411 & {0.2478}& 0.2718 & 0.3780  &0.6908& 0.9998 & 0.6915 & 0.8469 & 0.7579 & { \underline{0.2122}}& \textbf{0.1375} \\
& LPIPS $(\uparrow)$ & 0.1894 & \underline{0.5830} & 0.5754 & 0.6347  &0.2747& 0.0001 & 0.3223 & 0.1780 & 0.2631 & { 0.5821}& \textbf{0.5866} \\ \hline

\multirow{4}{*}{EMNIST} 
& MSE $(\uparrow)$ & 0.0003 & \underline{0.0633} & 0.0575 & 0.0057  &0.0017& 0.0000 & 0.0006 & 0.0003 & 0.0006 & { 0.0968}& \textbf{0.1429} \\
& PSNR $(\downarrow)$ & 36.8783 & {12.1025}& 12.5235 & 23.2789  &40.8229& inf & 35.7690 & 37.0652 & 33.6131 & {\underline{10.3058}}& \textbf{8.5792} \\
& SSIM $(\downarrow)$ & 0.9516 & {0.5376}& 0.5522 & 0.8178  &0.9833& 0.9968 & 0.9550 & 0.9553 & 0.9264 & {\underline{0.3084}}& \textbf{0.2025} \\
& LPIPS $(\uparrow)$ & 0.0111 & {0.5453}& 0.5310 & 0.1223  &0.0098& 0.0003 & 0.0135 & 0.0103 & 0.0176 & {\underline{0.6494}}& \textbf{0.6651} \\ \hline

HAR & MSE $(\uparrow)$ & 0.1953 & 0.2198 & 0.2907 & 0.2627  &0.2034& 0.000 & {0.2930}& 0.2493 & 0.2247 & { \underline{0.3561}} &\textbf{0.4156}\\ \hline
KWS & MSE $(\uparrow)$ & 0.0978 & 0.1286 & 0.1542 & 0.1129  &0.1194& 0.000 & \underline{0.1638} & 0.1385 & 0.1068 & { 0.1634} &  \textbf{0.1676}  \\ \hline \toprule
\end{tabular}
}
\end{table*}

\noindent \textbf{Datasets and Models:} We evaluate the effectiveness of our defense using the following datasets and applications.

\begin{itemize}
    \item  \textbf{Image Classification (IC)}. We consider two image classification datasets: the extended MNIST (EMNIST)~\cite{cohen2017emnist} with $28 \times 28$ grey-scale handwritten letters collected from different writers with imbalanced class distributions and CIFAR-10~\cite{krizhevsky73cifar} with $32 \times 32$ color images. 
    We randomly select 196 writers from the EMNIST dataset and assign each writer's data samples to a client. 
    For CIFAR-10, we create 100 class-imbalanced clients by splitting the data samples of each class based on proportions sampled from a Dirichlet distribution with $\alpha=0.5$.
    We denote the two applications as IC-EMNIST and IC-CIFAR10. 
    
    \item  \textbf{Human Activity Recognition (HAR)}. 
    HAR identifies daily activities like walking or sitting, leveraging sensor signals such as Inertial Measurement Unit (IMU) data. We adopt a public IMU dataset~\cite{anguita2013public} consisting of six daily activities 
    collected from 30 class-imbalanced participants. We assign each participant's data samples to a client. 

    \item \textbf{Keyword Spotting (KWS)}. 
    KWS captures specific commands using device microphones to enable voice-based human-computer interaction. We use the Google Speech Commands Dataset~\cite{warden2018speech}, which contains 105,829 one-second utterances of 35 keywords collected from different speakers. Each voice sample is processed into an $81 \times 40$ Mel-Frequency Cepstral Coefficients (MFCC) tensor.
    We randomly select 489 speakers that are class-imbalanced and assign each speaker's data samples to a client. 
\end{itemize}
For IC-CIFAR10, IC-EMNIST, and KWS, we adopt the ResNet-18 architecture. For HAR, we use the 1D ConvNet~\cite{kiranyaz20211d}. 
{ We also consider the larger-scale Vision Transformer (ViT)~\cite{dosovitskiy2020image} trained on ImageNet~\cite{russakovsky2015imagenet} in \sect\ref{subsec:defense-performance}.}
The total number of clients and the number of participating clients per communication round are summarized in Table~\ref{tab:num-client}. Note that the data splits in all the applications are class-imbalanced.

\noindent \textbf{Attack and Defense Baselines:}
We implement the widely considered IG attack as our primary evaluation attack due to its broad applicability across different model architectures and data types.
\sect\ref{subsec:defense-performance} also evaluates our defense using a recent strong GAN-based attack, ROG~\cite{yue2023gradient}, on high-resolution image data.
We consider { eight} representative defense baselines, including perturbation-based (DP~\cite{zhu2019deep}, Outpost~\cite{wang2023more}, CENSOR~\cite{zhang2025censor}, and PRECODE~\cite{scheliga2022precode}), pruning-based (Prune~\cite{zhu2019deep}, Soteria~\cite{sun2021soteria}, and DGP~\cite{xue2024revisiting}), { and compression-based ($p$FGD~\cite{palihawadana2023mitigating}) defenses.}
We implement adaptive attacks as defined in \sect\ref{sec:threat-model} and employ the same attack settings as in \sect\ref{sec:motivation-study} for all the defenses except for noise injection-based (i.e., DP and Outpost) { and compression-based (i.e., $p$FGD)} defenses. 
This is because noise injection-based defenses are theoretically proven effective against practical adaptive attacks in Appendix~\ref{subsec:analysis-gradient-perturb}, { while $p$FGD compresses the entire gradient space by discarding low-frequency components, similar to truncated SVD.}
However, \sect\ref{subsec:defense-performance} compares our defense with noise injection-based { and compression-based defenses} under a less practical adaptive adversary to demonstrate our defense’s robustness even under extreme threat conditions. 
We implement two DP baseline variants, DP-Gaussian and DP-Laplace, which employ Gaussian and Laplacian noises, respectively.

For IC-CIFAR10, IC-EMNIST, and KWS, we set the noise scale to 0.03. For HAR, we set the noise scale to 0.5. 
{ We follow~\cite{palihawadana2023mitigating} and set the pruning rate to 0.01 for $p$FGD.}
We randomly sample 128 input samples from each dataset to launch attacks. 
We set the local training batch size, number of epochs, and the number of steps all to be 1 and perform the attack in the first communication round. This setting is ideal for the adversary and most challenging for the defender, which is commonly adopted by existing defenses~\cite{wu2023learning, yue2023gradient,xue2024revisiting, scheliga2023dropout}.
By default, we set $\beta$ to be 0.3 in \textit{SVDefense} as our experiments in \sect\ref{sensitivity} show that this value achieves a favorable trade-off among defense, accuracy, and communication cost.

\noindent \textbf{Evaluation Metrics:} We use accuracy to evaluate classification performance. To evaluate defense effectiveness, we use metrics including the MSE, PSNR, structural similarity index measure (SSIM)~\cite{wang2004image}, and 
LPIPS.  
Note that we use all four metrics for image classification and only use MSE for the remaining applications, since PSNR, SSIM, and LPIPS are specific to image data.
Higher MSE and LPIPS and lower PSNR and SSIM indicate more effective defense performance.
To evaluate system overhead, we define the \textit{normalized on-device latency} and \textit{communication cost reduction}. 
The normalized on-device latency is defined as the ratio between the total on-device local training time with and without defense. A value of 1.0 for this metric indicates no additional computational overhead compared with the baseline without defense.
The communication cost reduction measures the percentage decrease in total communication latency (including both uploading and downloading latency) achieved by the defense relative to the baseline without defense. 

\noindent \textbf{Implementation Details:}
To validate the practicality of \textit{SVDefense}, we implement a real-world FL testbed { as shown in Appendix~\ref{app:visualization-reconstructed}}.
The testbed contains heterogeneous embedded platforms, including two NVIDIA Jetson TX2, two NVIDIA Jetson Nano, and six Raspberry Pi 4, as client devices. The server is equipped with an AMD EPYC 7543@ 3.7GHz, 256G RAM, and 4 RTX A5000 GPUs. We use TL-SG116 to connect the server and client devices. Since the number of clients participating in each communication round may exceed the number of available devices, we randomly assign a device for each client to train its local model. Each device trains one client model at a time due to resource constraints.

\subsection{Accuracy}
\label{acc}

Fig.~\ref{fig_acc} presents the classification accuracy across communication rounds in the presence of different defenses and in the absence of defense, indicated by ``None''. From the results, we can see that \textit{SVDefense} performs better than the undefended baseline ``None'' and the other defense methods across all the applications. 
As expected, DP-based defenses degrade the model utility. Specifically, DP-Laplace achieves a final accuracy of 51.0\% and DP-Gaussian achieves 53.5\% in KWS, while the ``None'' baseline achieves a final accuracy of 78.0\% and \textit{SVDefense} achieves 79.9\%, respectively.
Outpost achieves an accuracy similar to ``None'' by adaptively adjusting perturbations to preserve more information during the FL training. However, it has degraded defense performance, as shown in \sect\ref{subsec:defense-performance}. 
CENSOR maintains an accuracy close to ``None'' since its defense operations are only activated during the initial few training epochs~\cite{zhang2025censor}.
When the defense is activated for all the training epochs, CENSOR reduces the final accuracy by 14\%, 8\%, 18\%, and 9\% for IC-CIFAR10, IC-EMNIST, HAR, and KWS, respectively, compared with ``None''.
PRECODE has degraded accuracy in all the applications because it needs to sample a random vector in each training step, leading to extra noise.
Note that Soteria is omitted from the accuracy comparison as its layer-wise defense operations become computationally intractable when applied in each communication round~\cite{wang2023more}.
Prune achieves a lower accuracy due to its aggressive gradient pruning that discards potentially important update information. 
Although DGP performs well in IC applications, its accuracy fluctuates in HAR and KWS, potentially due to the per-step pruning of large gradients that may contain useful information. { $p$FGD has degraded accuracy because it applies the Discrete Cosine Transform to the gradients and directly sets the coefficients of the low-frequency components to zero, which may discard critical gradient direction information.}

\subsection{Defense Performance}
\label{subsec:defense-performance}

Table~\ref{tab:defense} presents the defense performance of different methods, with the best and second-best results highlighted in bold and underlined, respectively. Note that the PSNR is computed by dividing the MSE. Thus, ``inf'' values in the table indicate near-zero MSE values.
We can see from the table that \textit{SVDefense} achieves the best defense performance in all the applications, compared with all the baselines.
Although DP provides theoretical privacy guarantees, it significantly impacts the model utility, as shown in Fig.~\ref{fig_acc}. 
In comparison, \textit{SVDefense} achieves strong defense performance without compromising classification accuracy, due to the effectiveness of
the Channel-Wise Weighted Approximation and Layer-Wise Weighted Aggregation mechanisms in our design.
{ 
Examples of the reconstructed images under different defenses are provided in Appendix~\ref{app:visualization-reconstructed}. 
}

\begin{table}[t]

\caption{Comparison of Defense Effectiveness Across Different Defense Methods Under Adaptive LTI attack~\cite{wu2023learning}.} \label{tab:LTI}


\centering

\begin{tabular}{cccccc}

\hline

\toprule

Metric & DP-Gau & DP-Lap & Outpost & { $p$FGD} &\textit{SVDefense} \\ \hline

MSE $(\uparrow)$ & 0.0292& \underline{0.0315}& 0.0220 & { 0.0197} & \textbf{0.0469}\\

PSNR $(\downarrow)$ & 15.8955& \underline{15.5623}& 17.1465 & { 17.6388} & \textbf{14.3392}\\

SSIM $(\downarrow)$ & 0.2547& \underline{0.2356} & 0.3369 & { 0.3672} &{\textbf{0.1509}}\\

LPIPS $(\uparrow)$ & 0.5744& \underline{0.5834}& 0.5487 & { 0.5362} &\textbf{0.6521} \\ \hline

\toprule

\end{tabular}

\end{table}

\begin{table*}[t!]
\caption{Comparison of Defense Effectiveness Across Different Defense Methods on High-resolution ImageNet with LeNet~\cite{lecun1998gradient}.} \label{tab:imagenet}
\centering
\begin{tabular}{cccccccccccc}
\hline
\toprule
 Metric& None&  DP-Gau&  DP-Lap &Outpost
&   CENSOR&PRECODE
&  Prune
&  DGP
& { $p$FGD}
&  \textit{SVDefense}\\ \hline
 MSE $(\uparrow)$&  0.0220&  {0.0381}&  0.0369&0.0273
&   0.0289&0.0029
&  0.0265
&  0.0247
&  {\underline{0.0564}}
&  \textbf{0.0904}\\
 PSNR $(\downarrow)$&  17.2417&  {14.6213}&  14.6889&16.3300
&   16.3367&28.6856
&  16.4031
&  16.8004
&  {\underline{13.4950}}
&  \textbf{10.9315}\\
 SSIM $(\downarrow)$&   0.5090&  0.2613&  \underline{0.2446}  &0.4253
&   0.4162&0.9287
&  0.4280
&  0.4952
&  {0.4490}
&  \textbf{0.1128}\\
 LPIPS $(\uparrow)$&  0.4313&  0.6175&  \underline{0.6242} &0.4908&   0.5163&0.0236&  0.5053&  0.4498&{ 0.5343} & \textbf{0.7004}\\ \hline \toprule
\end{tabular}
\end{table*}

\begin{table*}[t!] 
\caption{{ Comparison of Defense Effectiveness Across Different Defense Methods on High-resolution ImageNet with ViT~\cite{dosovitskiy2020image}.}} \label{tab:vit}
\centering
\begin{tabular}{ccccccccccc}
\hline
\toprule
 Metric& None&  DP-Gau&  DP-Lap &Outpost
&   CENSOR&PRECODE
&  Prune
&  DGP
& $p$FGD
&  \textit{SVDefense}\\ \hline
 MSE $(\uparrow)$&  0.0817 &  0.0796&  0.0794&0.0848&   0.0874&  0.0834
&  \underline{0.1109}
&  0.0859
&  0.0985
&  \textbf{0.1287}\\
 PSNR $(\downarrow)$&  11.4922 &  11.6943&  11.7168&11.3691&   11.1634&  11.4756
&  \underline{9.9646}
&  11.2950
&  10.5598
&  \textbf{9.2805}\\
 SSIM $(\downarrow)$&   0.4852 &  0.2795&  0.2704&0.1925&   0.4342&  0.4847
&  0.3194
&  0.4586
&  \underline{0.1821}
&  \textbf{0.0494}\\
 LPIPS $(\uparrow)$&  0.3528 & 0.6552&  0.6739&0.6939&   0.3793& 0.3536 &  0.5627 &  0.3834 & \underline{0.6631} &\textbf{0.7473}\\ \hline \toprule
\end{tabular}
\end{table*}

\begin{figure}[t!]
  \centering
  \subfloat[SVD]{
    \centering
    \includegraphics[width=0.464\columnwidth]{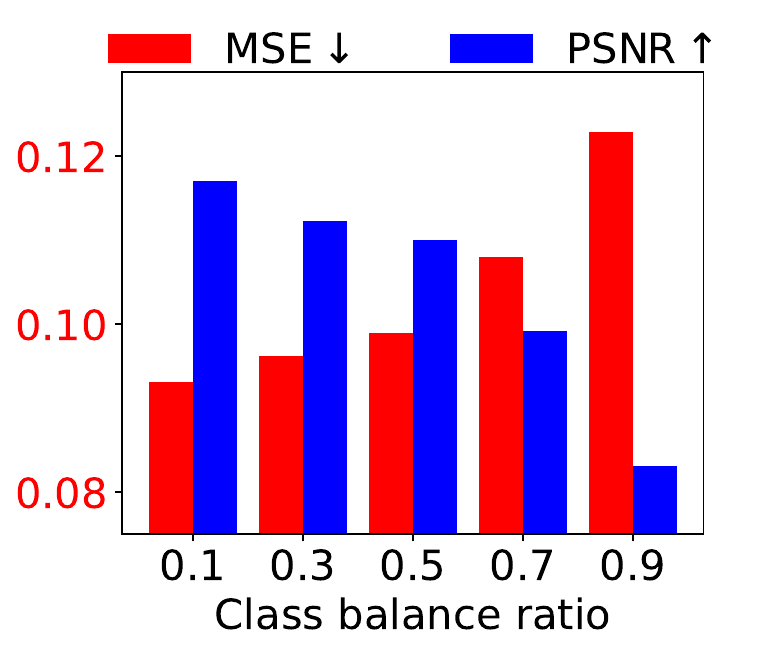}
    }
  \hfill
  \centering
  \subfloat[SVD\_S]{
    \centering
    \includegraphics[width=0.46\columnwidth]{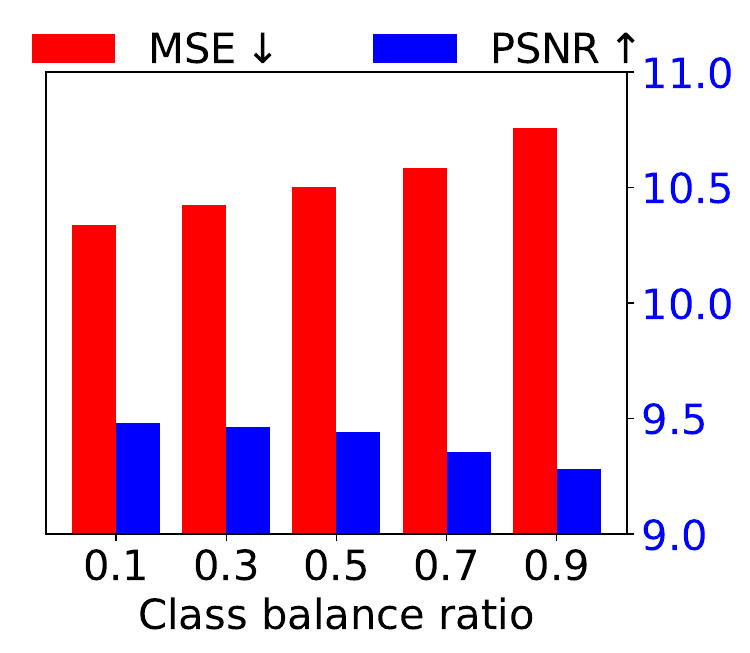} 
  }
  \caption{
  Impact of Self-Adaptive Energy Threshold on defense performance under class imbalance. Lower MSE and higher PSNR indicate stronger attack effectiveness. { $\mathcal{T} = 0.8$ is fixed for ``SVD''; $\beta$ for ``SVD\_S'' is chosen to match the defense performance of ``SVD'' at class balance ratio of 0.9.}}
  \label{fig:defense_adaptive}  %
  \end{figure}
\noindent \textbf{Adaptive Adversary against \textit{SVDefense}.} 
As analyzed in \sect\ref{sec:motivation-study}, \textit{SVDefense}
is invulnerable to adaptive attack operations like identifying modified gradient components or recovering random variables used by the defender.
Nevertheless, to further evaluate \textit{SVDefense}'s performance in extreme situations, we simulate a powerful adaptive attacker based on the LTI attack~\cite{wu2023learning}.
Specifically, we assume that the attacker has obtained a surrogate training dataset that follows the same distribution as the client's local data. The attacker can then train a surrogate local model on this dataset, use the model to generate input-gradient pairs, and apply \textit{SVDefense} to process these gradients. The attacker then trains a neural network to learn the mapping from the approximated gradients to the input samples. We can follow a similar procedure to train the gradient inversion models for the DP, Outpost, and { $p$FGD} baselines. During the attack phase, the gradient inversion model is applied to a victim client's defended gradients in order to attempt data reconstruction. Table~\ref{tab:LTI} presents the defense performance on CIFAR-10. 
From the results, we can observe that 
\textit{SVDefense} achieves the best defense performance compared with the DP baselines (i.e., DP-Gaussian and DP-Laplace both with a noise scale of 0.1), Outpost, and { $p$FGD}. 
However, the assumption that the adversary can obtain the distribution of a client's real data can be impractically strong and work against the premise of GIAs. This is because in real-world FL scenarios, the private data distribution is precisely what a client aims to protect. If the attacker already had access to similarly distributed data, there would be little motivation to perform GIAs in the first place.


\noindent \textbf{\textit{SVDefense} Effectiveness for High-Resolution Images { and on Large-Scale Model}.} 
We evaluate our defense on ImageNet, a large-scale dataset consisting of 1,000 classes of high-resolution ($224 \times 224$) color images. We train a LeNet~\cite{lecun1998gradient} on ImageNet and implement a recent strong GAN-based attack, ROG. We follow the same evaluation setup as described in \sect\ref{subsec:eval-setup} except that we set the training batch size to 16 and the noise scale of the DP baselines to 0.1. 
As shown in Table~\ref{tab:imagenet}, \textit{SVDefense} achieves superior defense performance compared with all the baselines. Note that Soteria is excluded from this comparison due to its computational infeasibility for high-dimensional data. { We then evaluate \textit{SVDefense}'s performance on the large-scale ViT model using the IG attack. As shown in Table~\ref{tab:vit}, \textit{SVDefense} still outperforms all the baselines.}


\subsection{Impact of Self-Adaptive Energy Threshold under Class Imbalance}
This section evaluates the impact of the Self-Adaptive Energy Threshold on the defense performance of truncated SVD under class imbalance. We follow the method described in \sect\ref{subsec:challenges} to simulate varying degrees of class imbalance on CIFAR-10. 
Fig.~\ref{fig:defense_adaptive} illustrates the defense performance of truncated SVD with and without Self-Adaptive Energy Threshold, denoted by ``SVD'' and ``SVD\_S'', respectively, under varying degrees of class imbalance. For a fair comparison, we fix the energy threshold $\mathcal{T}$ to be 0.8 for ``SVD'' and select a $\beta$ for ``SVD\_S'' such that its defense performance matches that of the ``SVD'' when the class balance ratio is 0.9. We can observe that the defense performance of ``SVD'' deteriorates as the class balance ratio decreases. In comparison, ``SVD\_S'' effectively adapts to varying degrees of class imbalance and maintains more stable defense performance. The defense performance for ``SVD\_S'' is always better than that of  ``SVD''. This is because class-imbalanced inputs produce gradients with a more skewed distribution of squared singular values, leading the Self-Adaptive Energy Threshold to adaptively derive a lower energy threshold that provides stronger protection.
{
\subsection{Impact of Channel-Wise Weighted Approximation}
\label{subsec:exp-channel-wise}

\begin{figure}[t!]
    \centering

    \subfloat[Classification accuracy vs. energy threshold.]{
        \includegraphics[width=.8\columnwidth]{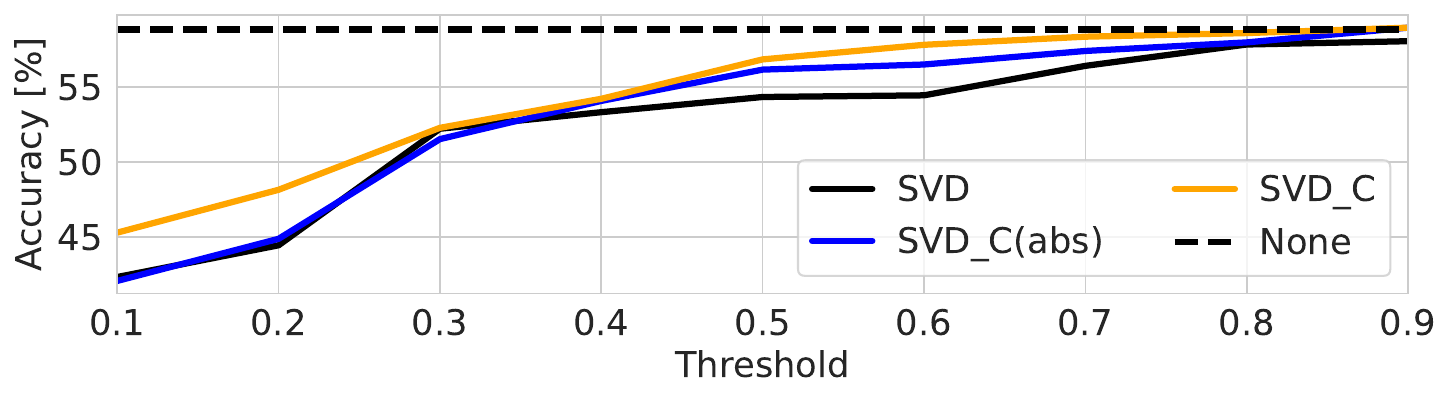}
        \label{fig:ablation_accuracy_T}
    } \\
    \subfloat[Defense performance vs. energy threshold.]{
        \includegraphics[width=.8\columnwidth]{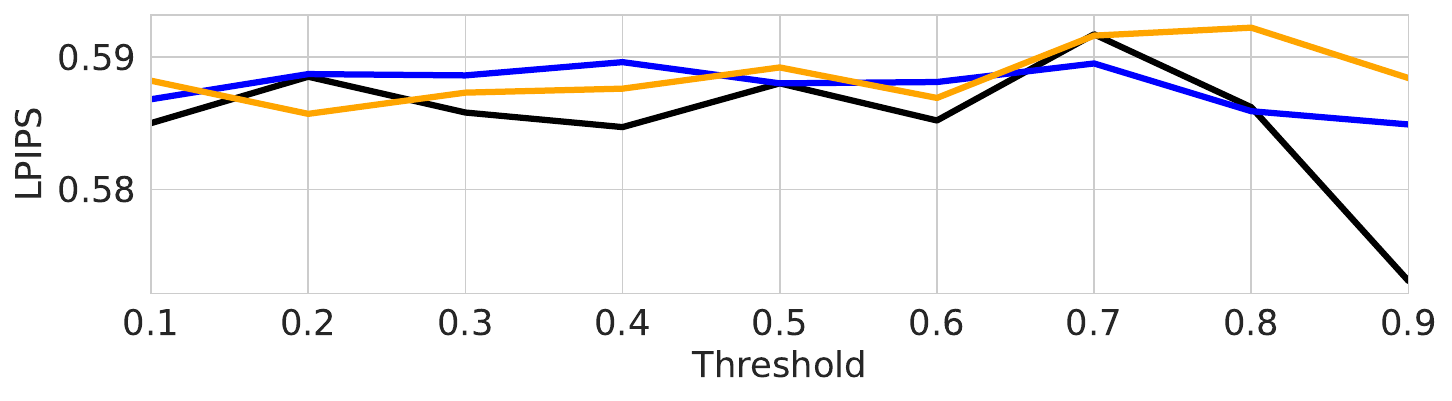}
        \label{fig:ablation_lpips_T}
    }
    \caption{Impact of varying energy threshold on accuracy and defense performance for SVD\_C.}
    \label{fig:ablation_performance_T}
\end{figure}

This section compares the original truncated SVD (``SVD'') and two variants of truncated SVD with Channel-Wise Weighted Approximation, i.e., ``SVD\_C'' that uses squared gradients as weights and ``SVD\_C(abs)'' that uses absolute values of gradients as weights, on CIFAR-10.
Fig.~\ref{fig:ablation_accuracy_T} shows the classification accuracy when varying the energy threshold $\mathcal{T}$ from 0.1 to 0.9 with a step size of 0.1. We can observe that the accuracy increases with the threshold, and that ``SVD\_C'' outperforms both ``SVD\_C(abs)'' and ``SVD''. 
Fig.~\ref{fig:ablation_lpips_T} shows the defense performance. We can see that, when $\mathcal{T} < 0.7$, the LPIPS value fluctuates and all the methods perform similarly. When $\mathcal{T}$ is greater than 0.7, ``SVD\_C'' outperforms both ``SVD\_C(abs)'' and ``SVD''. These results demonstrate the effectiveness of the Channel-Wise Weighted Approximation module in enhancing both the accuracy and defense performance. They also show that, compared with absolute values of gradients as weights, square gradients as weights better emphasize larger singular components, which contributes to improved classification and defense performance.
}

\begin{figure}[t!]
    \centering

    \subfloat[Classification accuracy vs. $\beta$.]{
        \includegraphics[width=.8\columnwidth]{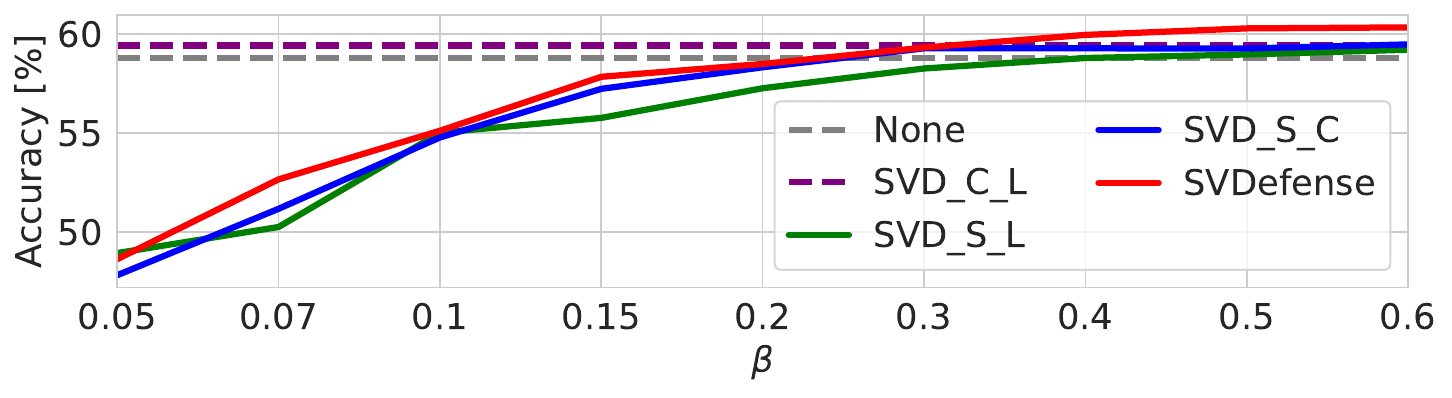}
        \label{fig:ablation_accuracy}
    } \\
    \subfloat[Defense performance vs. $\beta$.]{
        \includegraphics[width=.8\columnwidth]{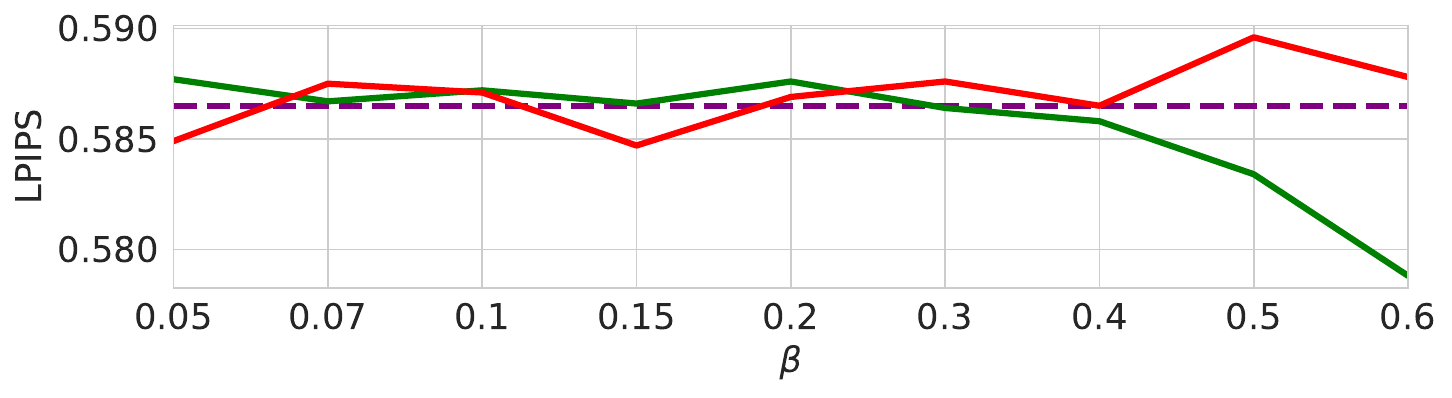}
        \label{fig:ablation_lpips}
    }
    \caption{Impact of varying $\beta$ on accuracy and defense performance for \textit{SVDefense}.}
    \label{fig:ablation_performance}
\end{figure}

\begin{figure}[t!]
    \centering
    \begin{minipage}{0.44\columnwidth} 
        \centering
        \includegraphics[width=\linewidth]{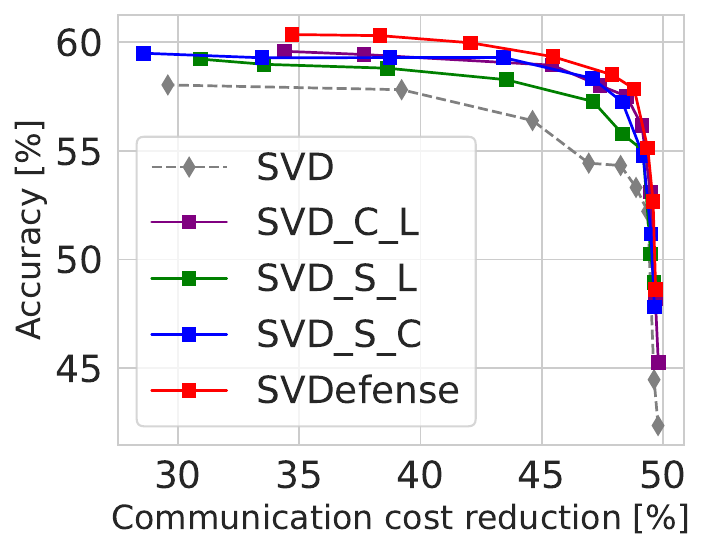} 
        \caption{Classification accuracy vs. communication cost reduction.}
        \label{fig:acc_com}
    \end{minipage}
    \hfill
    \begin{minipage}{0.53\columnwidth}
        \centering
        \includegraphics[width=\linewidth]{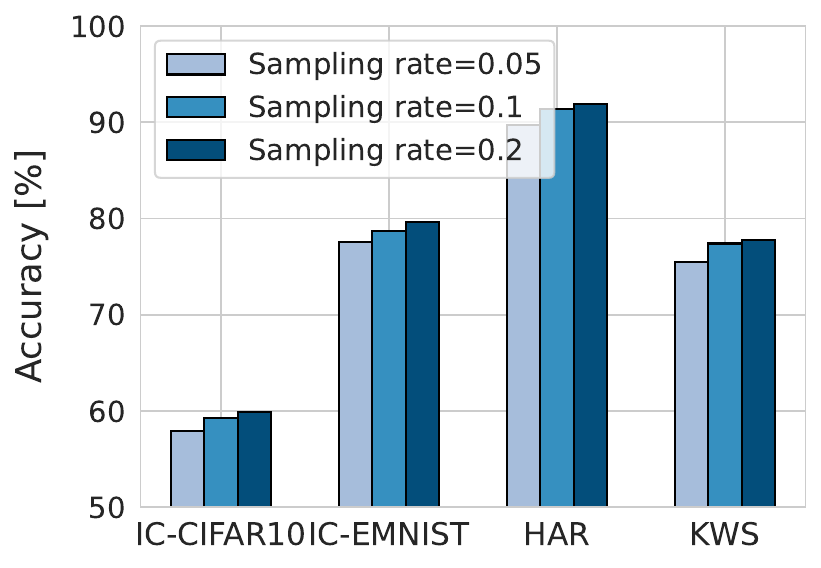}
        \caption{Impact of client sampling rate on \textit{SVDefense} classification accuracy.}
        \label{fraction_acc}
    \end{minipage}
\end{figure}

\subsection{Ablation Study}
\label{subsec:ablation-study}

We evaluate three key components of \textit{SVDefense}, namely Self-Adaptive Energy Threshold, Channel-Wise Weighted Approximation, and Layer-Wise Weighted Aggregation, on CIFAR-10. We denote different variants as ``SVD\_\{\}\_\{\}'', where each placeholder in brackets contains the component's initial letter if included. For example, ``SVD\_S\_C'' represents \textit{SVDefense} without the Layer-Wise Weighted Aggregation.
``SVDefense'' denotes our full proposed method.

Fig.~\ref{fig:ablation_performance} presents the accuracy and defense performance of different variants. For the variant without the Self-Adaptive Energy Threshold, we set a fixed energy threshold $\mathcal{T}$ of 0.8, which achieves a balanced accuracy and defense performance based on empirical results. For variants with the Self-Adaptive Energy Threshold, we vary the sensitivity parameter $\beta$ within \{0.05, 0.07, 0.1, 0.15, 0.2, 0.3, 0.4, 0.5, 0.6\}. Since the undefended baseline ``None'' and ``SVD\_C\_L'' are not affected by the variation of $\beta$, their performance is represented by horizontal lines in Fig.~\ref{fig:ablation_performance}.
The Layer-Wise Weighted Aggregation aims to improve the global model's utility and does not affect the defense performance. Therefore, in Fig.~\ref{fig:ablation_lpips}, we only present ``SVDefense'' (equivalent to ``SVD\_S\_C''), ``SVD\_S\_L'' (equivalent to ``SVD\_S''), and ``SVD\_C\_L'' (equivalent to ``SVD\_C''). From Fig.~\ref{fig:ablation_accuracy}, the complete version of \textit{SVDefense} achieves the best accuracy compared with all the other variants. From Fig.~\ref{fig:ablation_lpips}, the defense performance of all the methods is similar when $\beta < 0.3$. When $\beta \geq 0.3$, \textit{SVDefense} outperforms the other variants. In conclusion, \textit{SVDefense} that combines all three components achieves the best accuracy and defense performance when setting $\beta$ at appropriate values. This is because the Self-Adaptive Energy Threshold and Channel-Wise Weighted Approximation effectively suppress sensitive information leakage under class imbalance while preserving more information critical for the model training. The Layer-Wise Weighted Aggregation further enhances the global model accuracy.
Fig.~\ref{fig:acc_com} shows the classification accuracy versus the communication cost reduction of the different variants.
For the variants without Self-Adaptive Energy Threshold, we vary $\mathcal{T}$ from 0.1 to 0.9 with a step size of 0.1. For the variants with Self-Adaptive Energy Threshold, we vary $\beta$ to be \{0.05, 0.07, 0.1, 0.15, 0.2, 0.3, 0.4, 0.5, 0.6\}. We can see that \textit{SVDefense} achieves better accuracy and communication efficiency, compared with all the other variants. 
This validates the effectiveness of the three key components in balancing accuracy and communication efficiency.

\subsection{Sensitivity Analysis}
\label{sensitivity}

This section evaluates the impact of varying different parameters on \textit{SVDefense}'s performance using the CIFAR-10 dataset.

\begin{figure*}[t!]
\centering
\includegraphics[width=\textwidth]{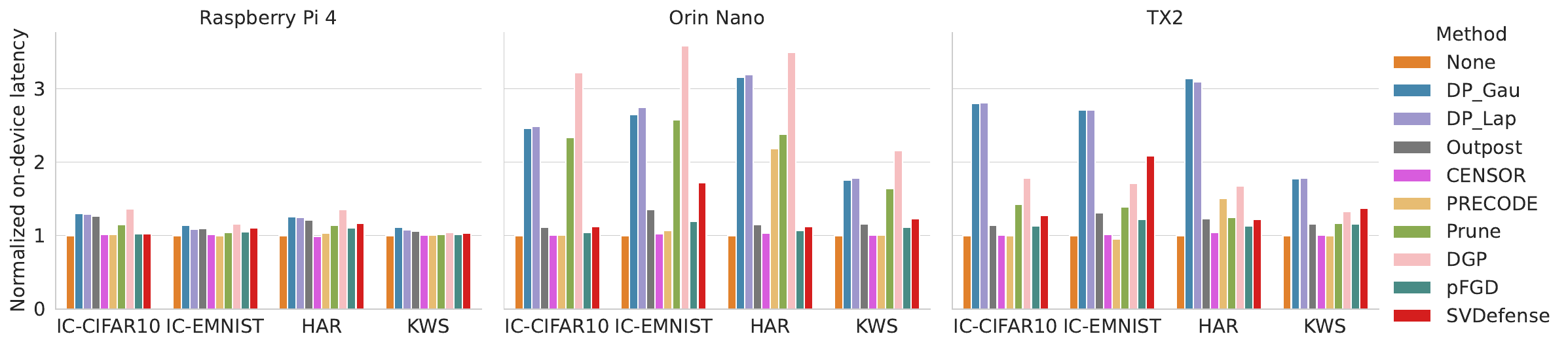}
\centering
\caption{Comparison of normalized on-device latency across different defense methods on three embedded platforms.}
\label{computation_cost}
\end{figure*}

\noindent \textbf{Sensitivity Parameter $\beta$.}
First, we analyze the impact of varying the sensitivity parameter $\beta$ on \textit{SVDefense}'s performance. We vary the value of $\beta$ to be \{0.05, 0.07, 0.1, 0.15, 0.2, 0.3, 0.4, 0.5, 0.6\}. The red line in Fig.~\ref{fig:ablation_accuracy} shows the classification accuracy of \textit{SVDefense} across different $\beta$ values. 
The result shows that the accuracy increases with $\beta$. This is because for the same entropy value, a larger value of $\beta$ leads to a higher energy threshold $\mathcal{T}$, retaining more gradient information. This improvement in accuracy plateaus around $\beta \approx 0.3$, suggesting that additional gradient information beyond this point contributes minimally to the classification performance.
The red line in Fig. \ref{fig:ablation_lpips} illustrates the defense performance of \textit{SVDefense} across different $\beta$ values. The performance fluctuates because \textit{SVDefense} tends to retain larger gradients for higher accuracy, which can also be primarily exploited for data reconstruction~\cite{xue2024revisiting}. 
However, by perturbing the gradients with channel-wise weights and aggregating the local models with layer-wise weights, \textit{SVDefense} can effectively improve both the accuracy and defense performance.
The red line in Fig.~\ref{fig:acc_com} illustrates the trade-off between the accuracy and communication cost reduction of \textit{SVDefense} across different $\beta$ values. The result shows that the accuracy decreases as the communication cost reduction rate increases. \textit{SVDefense} achieves a favorable trade-off when $\beta \in$ \{0.3,0.2,0.15\}. In summary,
\textit{SVDefense} achieves an optimal balance at $\beta \approx 0.3$, where it maintains strong classification performance, robust privacy protection, and high communication efficiency.
Similar sensitivity analysis can be applied to determine the optimal value of $\beta$ for other applications, considering their respective requirements for model accuracy, privacy protection, and communication cost. 


\noindent \textbf{Number of Participating Clients.} 
We analyze the impact of the per-round client sampling rate on the classification accuracy of \textit{SVDefense}.
We vary the client sampling rate $f \in$ \{0.05,0.1,0.2\}, where $f$ represents the ratio of sampled clients in each communication round.
As shown in Fig.~\ref{fraction_acc}, increasing the client sampling rate yields modest accuracy improvements. For example, for IC-CIFAR10, the accuracy increases by 1.97\% when $f$ increases from 0.05 to 0.2. However, this marginal performance gain comes with 4x higher client-server communication costs. The results suggest that moderate client sampling rates can achieve a good model performance while maintaining communication efficiency.

\subsection{System Overhead}


\begin{table}[t!]
\centering
\caption{Absolute On-device Latency (Seconds) of ``None''.}
\begin{tabular}{lcccc}
\toprule
Device & IC-CIFAR10 & IC-EMNIST & HAR & KWS \\
\midrule
RPi        & 130.1& 31.3& 2.1& 62.5\\
Orin Nano   & 3.6& 0.8& 0.2& 1.7\\
TX2         & 4.5& 1.1& 0.3& 2.6\\
\bottomrule
\end{tabular}
\label{tab:none_latency}
\end{table}

\begin{table}[t!]
\caption{Communication Cost Reduction (\%) for \textit{SVDefense}.}
\label{communication_cost}
    \centering
    \resizebox{\linewidth}{!}{
    \begin{tabular}{ccccc}  
        \hline
        \toprule
        \textbf{Application} & IC-CIFAR10 & IC-EMNIST & HAR & KWS \\
        \hline
        \makecell{\textit{Comm. cost} \\ \textit{reduction (\%)}} & 42.0 & 30.3 & 48.6 & 23.7 \\
        \hline
        \toprule
    \end{tabular}
    }
\end{table}

\noindent \textbf{Normalized On-Device Latency.} 
Fig.~\ref{computation_cost} compares the normalized on-device latency of different defenses on three embedded platforms. The absolute on-device latency of the undefended baseline ``None'' over one epoch can be found in Table~\ref{tab:none_latency}.
First, on the Raspberry Pi 4, all the defenses incur minimal additional computation overhead. This is because the CPU-based model training on the Raspberry Pi 4 is time-intensive, making the defense operation time relatively insignificant.
Second, the DP-based defenses show a 2-3× slowdown compared with ``None'' on both the Orin Nano and TX2. This is because random sampling operations are costly on resource-constrained hardware.
Third, pruning-based defenses perform notably slower on Orin Nano due to their computation-intensive matrix operations like sorting.
Fourth, CENSOR and { $p$FGD} have negligible extra computational overhead but struggle to achieve a good balance between model utility and privacy protection.
Lastly, \textit{SVDefense} incurs limited additional on-device computational overhead on all the platforms across all the applications except IC-EMNIST. This is because \textit{SVDefense} is only applied once at the end of each communication round, reducing the overall computational cost. The higher extra computational cost for EMNIST is because the dataset is relatively simple. Thus, the total local training time becomes comparable to the SVD operation time. 
Note that Soteria is omitted from the latency comparison as its layer-wise defense operations applied in each training step become computationally infeasible on these embedded platforms~\cite{wang2023more}.

\textbf{Communication Cost Reduction.}
\textit{SVDefense} achieves communication cost reduction by decomposing and truncating the model updates into matrices with fewer parameters than the original updates. As shown in Table~\ref{communication_cost}, \textit{SVDefense} significantly reduces the communication costs in all the applications. 
Note that the other defense baselines have similar communication cost as ``None'' because these defenses do not have specific mechanisms for communication time reduction.

\section{Discussion}
\textit{SVDefense} can be potentially extended to deep learning architectures like recurrent neural networks (RNNs), graph neural networks (GNNs), and other emerging model architectures. For example, SVD can be applied to the recurrent layers in RNNs and the message-passing layers in GNNs. 
Doing so may require developing new SVD strategies that account for the unique characteristics of these architectures. 
Beyond the applications discussed in this paper, \textit{SVDefense} can be employed in other domains such as natural language processing~\cite{deng2021tag,balunovic2022lamp,feng2024uncovering} and medical image analysis.

\section{Conclusion}

This paper presents \textit{SVDefense}, a novel defense framework based on truncated SVD against adaptive GIAs in FL. Our framework introduces three key innovations: the Self-Adaptive Energy Threshold that adapts to client vulnerability, the Channel-Wise Weighted Approximation to enhance accuracy and defense performance, and the Layer-Wise Weighted Aggregation for effective aggregation under class imbalance. Extensive experiments demonstrate that \textit{SVDefense} outperforms existing defenses in both model accuracy and defense effectiveness. Furthermore, \textit{SVDefense} achieves practical computational cost and much reduced communication cost on a real-world FL testbed.

\section*{Acknowledgment}

We thank the anonymous reviewers for their valuable comments and suggestions.



\bibliographystyle{IEEEtran}
\bibliography{IEEEabrv,reference}

\appendix

\section{Appendix}



\subsection{Analysis of Noise Injection-based Defenses under EoT Attack}
\label{subsec:analysis-gradient-perturb}

In GIAs, the adversary optimizes the dummy input $\vec{x^{\prime}}$ by minimizing $\mathcal{D}(\vec{\nabla\Theta},\vec{\nabla\Theta^{\prime}})$, where $\vec{\nabla\Theta}$ and $\vec{\nabla\Theta^{\prime}}$ represent the ground truth and dummy gradients, respectively, and $\mathcal{D}$ represents the distance metric.
The distance metric $\mathcal{D}$ quantifies the distance between the two gradients. To simplify our analysis,
we consider $\mathcal{D}$ to be the Euclidean distance. 
Under defense, the GIA objective is:
\begin{equation}
    \begin{aligned}
        \mathrm{min}~\|\varphi(\vec{\nabla\Theta}) - \vec{\nabla\Theta^{\prime}}\|_{F},
    \end{aligned}
    \label{eqn:attack_obj}
\end{equation}
where $\varphi(\cdot)$ denotes the defense operation and $\|\cdot\|_F$ is the Frobenius norm.

Take DP-Gau \cite{zhu2019deep} as an example, $\varphi(\cdot)$ represents injecting noise $\vec{\eta}$ sampled from a Gaussian distribution $\mathcal{N}(0, \sigma^2)$ to input data. Then, the observed gradients can be denoted by $\varphi(\vec{\nabla\Theta}) = \vec{\nabla\Theta} + \vec{\eta}$ and the optimization objective of GIA under DP-Gau becomes: 
\begin{equation}
    \begin{aligned}
    \mathrm{min}~\|\vec{\nabla\Theta} + \vec{\eta} - \vec{\nabla\Theta^{\prime}}  \|_{F}. 
    \end{aligned}
    \label{eqn:dp-gau-attack-effectiveness}
\end{equation}
When an adaptive adversary applies the Expectation over Transformation (EoT)~\cite{zhang2025censor} by sampling $n$ times from the Gaussian distribution $\mathcal{N}(0, \sigma^2)$ to add noise to and average the perturbed input, the dummy gradients become $\vec{\nabla\Theta^{\prime}} + \vec{\eta^{\prime}}$, where $\vec{\eta^{\prime}} \sim \mathcal{N}(0, \frac{\sigma^2}{n})$. Consequently, the optimization objective becomes: 
\begin{equation}
    \begin{aligned}
    \mathrm{min}~\|\vec{\nabla\Theta} - \vec{\nabla\Theta^{\prime}}  + \vec{\eta} - \vec{\eta^{\prime}}\|_{F}, 
    \end{aligned}
    \label{eqn:dp-gau-eot-attack-effectiveness}
\end{equation}
where $\vec{\eta} - \vec{\eta^{\prime}}$ follows a Gaussian distribution $\mathcal{N}(0, \frac{n+1}{n}\sigma^2 )$.  
Given the objectives in Eqs.~\ref{eqn:dp-gau-attack-effectiveness} and~\ref{eqn:dp-gau-eot-attack-effectiveness}, the optimized dummy gradients $\vec{\nabla\Theta^{*}}$ is approximated as $\vec{\nabla\Theta + \vec{\eta}}$ and $\vec{\nabla\Theta} + \vec{\eta} - \vec{\eta^{\prime}}$, respectively. 
As theoretically proven in~\cite{xue2024revisiting}, the attack effectiveness of GIAs, which is characterized by the data reconstruction error, is lower bounded by the overall gradient error between the ground-truth and optimized dummy gradients, which is $\|\vec{\nabla\Theta - \vec{\nabla\Theta^{*}}}\|_F$. 
By applying the EoT operation, the data reconstruction error lower bound for DP-Gau changes from $\|\vec{\eta}\|_F$ to $\|\eta - \vec{\eta^{\prime}}\|_F$, with increased variance. Consequently, the attack becomes less effective with EoT operation.
A similar analysis can be done on Outpost that injects Gaussian noise in the gradients based on leakage risks.

For DP-Lap, the injected noise $\vec{\eta}$ follows Laplace distribution $Laplace(0,b)$ with mean 0 and variance $2b^2$. The noise $\vec{\eta^{\prime}}$ introduced in the dummy gradients when the adversary applies EoT follows a distribution with mean 0 and variance $\frac{2b^2}{n}$. Thus, $\vec{\eta} - \vec{\eta^{\prime}}$, follows a new distribution with mean 0 and variance  $\frac{(2+2n)b^2}{n}$, which is greater than the variance of $\vec{\eta}$, leading to an increased variance in the lower bound of the data reconstruction error. We can conclude that the EoT operation also deteriorates the attack effectiveness against DP-Lap.

In conclusion, under noise injection-based defenses including DP-Gau, Outpost, and DP-Lap, the adaptive attack operation of EoT increases the variance of the lower bound of data reconstruction error, leading to reduced attack effectiveness. 



\subsection{Effectiveness of Adaptive Attack under LRP and MQ}
\label{lrp}

\begin{table}[t]
    \caption{Defense Performance of LRP { and MQ} Under Non-adaptive and Adaptive GIAs.}
    \centering
    \begin{tabular}{|c|c|c|c|c|}
    \hline
    \multirow{2}{*}{\diagbox{Defense}{Metric}} & \multicolumn{2}{c|}{Non-adaptive} & \multicolumn{2}{c|}{Adaptive} \\
      
    \cline{2-5}
     & PSNR & LPIPS & PSNR & LPIPS \\
     \hline
    LRP \cite{wan2023enhancing} & { 12.8634} & { 0.4995} & { 31.2368} & { 0.1459} \\
    \hline
    { MQ \cite{ovi2023mixed} } & { 4.3438} & { 0.7567} & { 45.7481} & { 0.0032} \\
    \hline
    \end{tabular}
    
    \label{tab:lrp}
\end{table}
This section evaluates the defense performance of LRP~\cite{wan2023enhancing} { and MQ~\cite{ovi2023mixed}} under both non-adaptive and adaptive attackers. 

LRP defends against GIAs by assigning randomly sampled learning rates to clients, concealing them from the attacker. In our experiments, we set the client learning rate to be 0.1. 
For the non-adaptive attack, we implement the DLG attack \cite{zhu2019deep}, as LRP has been shown to be resistant to DLG but vulnerable to the IG attack. We then configure the adversary’s dummy learning rate to be 0.2, simulating the LRP defense. For the adaptive attack against LRP, we initialize a trainable dummy learning rate and optimize it together with the dummy input during the attack optimization process. 

{
MQ defends against GIAs by hiding the gradient range information. For the non-adaptive attack, we implement the IG attack and directly use quantized gradients for input reconstruction. For the adaptive attack, similar to LRP, the adversary can initialize trainable dummy minimum and maximum gradient vectors, which are jointly optimized together with the dummy input during attack optimization process. The recovered hidden gradient range can be used to dequantize the gradients.

Table~\ref{tab:lrp} presents the defense performance of LRP and MQ. The results show that the adaptive attacks significantly weaken the defense effectiveness of both methods.
}

\subsection{Impact of Energy Threshold on Attack Effectiveness}
\label{threshold_mse}

\begin{figure}[t!]
\centering
\includegraphics[width=3.3in]{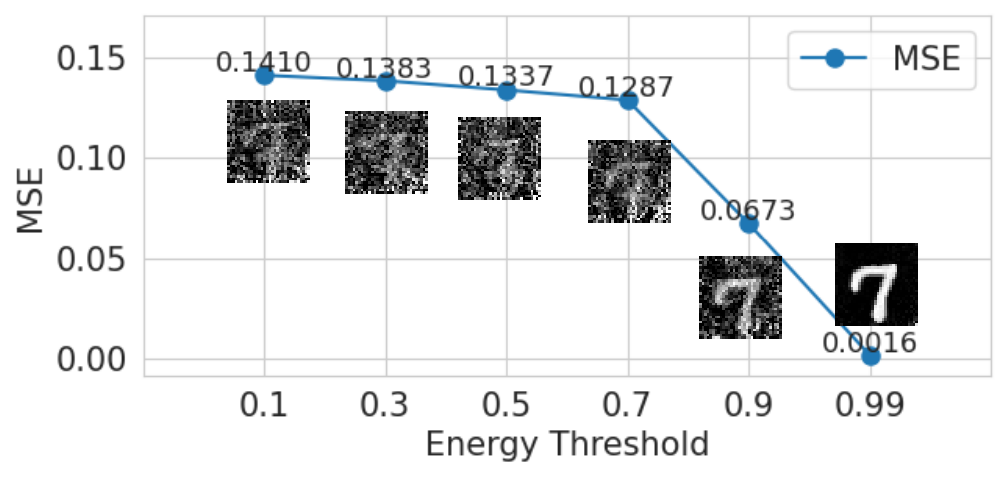}
\centering
\caption{The attack performance under different energy thresholds.}
\label{mse_threshold_7}
\end{figure}

\begin{figure}[t!]
\centering
\includegraphics[width=.6\columnwidth]{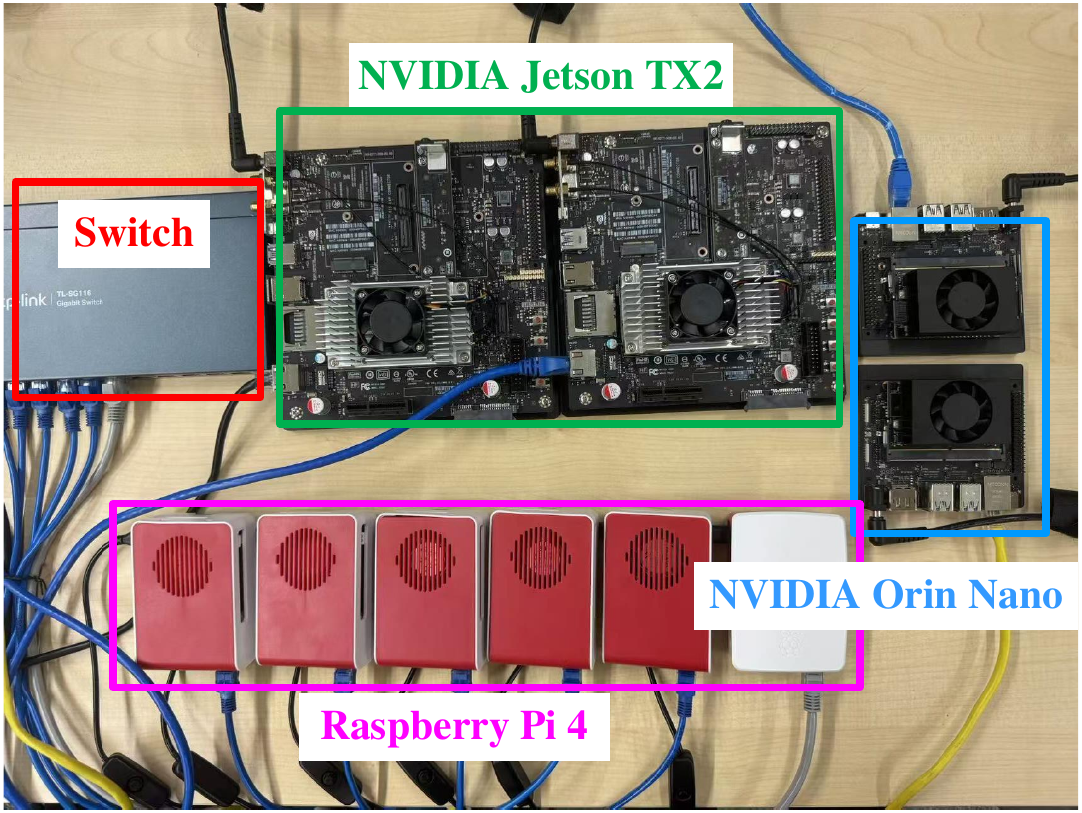}
\centering
\caption{An illustration of our federated learning testbed with various embedded platforms.}
\label{system}
\end{figure}

Fig.~\ref{mse_threshold_7} illustrates the effectiveness of IG attack under truncated SVD with varying energy thresholds using an image from the MNIST dataset. As the energy threshold increases, the MSE decreases and image reconstruction quality also increases. This is because truncated SVD with higher energy threshold retains more singular values and corresponding vectors, preserving more information for data reconstruction. This exemplifies that clients with smaller energy thresholds during SVD truncation receive stronger protection against GIAs.

\begin{figure*}[t!]
\centering
\includegraphics[width=.9\textwidth]{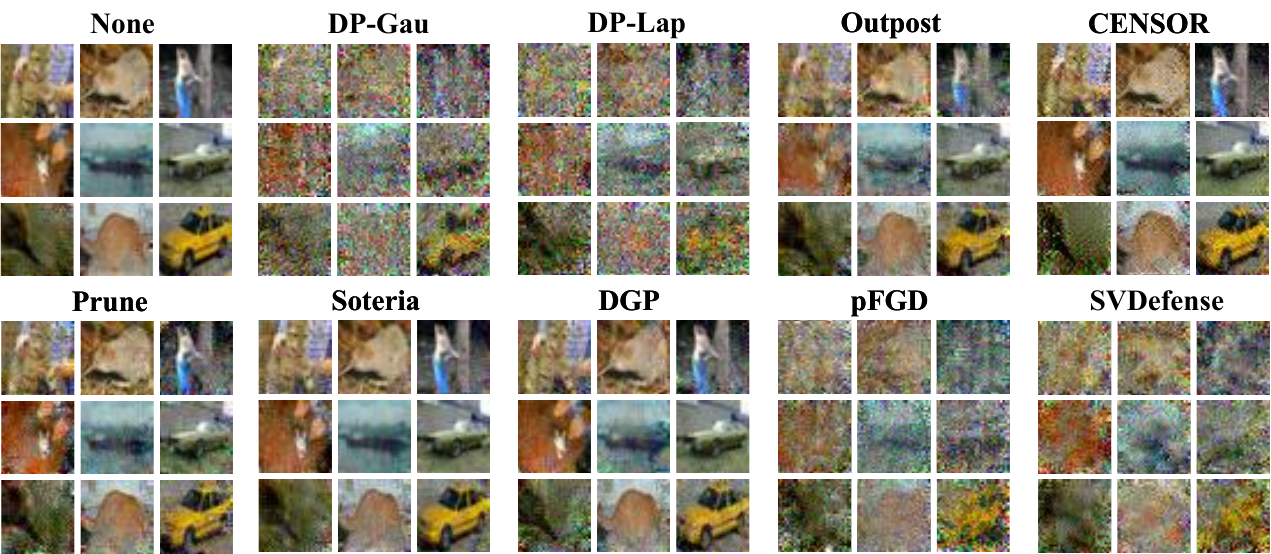}
\caption{Visual examples of reconstructed inputs on CIFAR-10.}
\label{fig:CIFAR10_visual}
\end{figure*}

\begin{figure*}[t!]
\centering
\includegraphics[width=.9\textwidth]{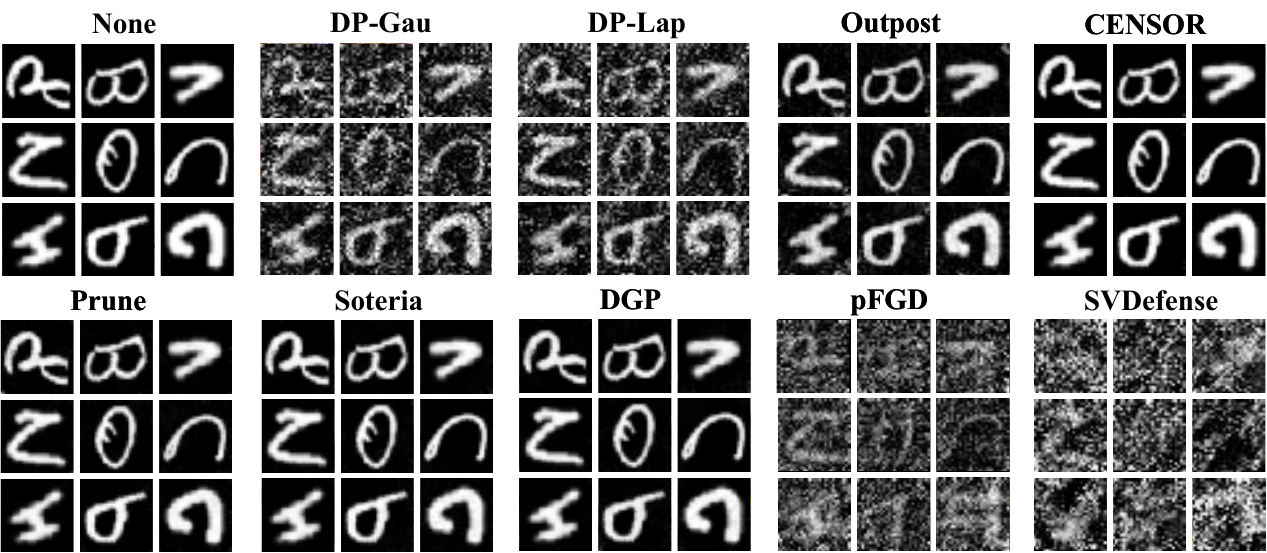}
\caption{Visual examples of reconstructed inputs on EMNIST.}
\label{fig:MNIST_visual}
\end{figure*}

\subsection{Theoretical Analysis of truncated SVD with Channel-Wise Weighted Approximation}
\label{app:proof-defense-performance}


\textbf{Theorem 1}. \textit{For any $(\varepsilon,\delta)$-passive attack $\mathcal{A}$, Under the presence of truncated SVD, it will degenerate to $(\varepsilon+\sqrt{\gamma_1} \|\vec{\nabla \Theta} \|_F,\delta)$, where $\gamma_1 = 1 -\mathcal{T}$. Under the presence of truncated SVD with Channel-Wise Weighted Approximation, it will degenerate to $(\varepsilon+\sqrt{\gamma_2} \|\vec{\nabla \Theta} \|_F,\delta)$, where $\gamma_2 = (\frac{\sigma_{max}(\vec{I})}{\sigma_{min}(\vec{I})})^2 (1-\mathcal{T})$, $\sigma_{max}(\cdot)$ and $\sigma_{min}(\cdot)$ mean the maximum and minimum singular value of the input matrix.}

\textit{Poof.} By definition, an $(\varepsilon,\delta)$-passive attack allows the attacker to achieve:
\begin{equation}
\label{equ:def}
\begin{aligned}
\mathbb{E} \|\vec{\nabla\Theta} -\vec{\nabla\Theta^{*}} \|_F \leq \varepsilon,
\end{aligned}
\end{equation}
where $\vec{\nabla\Theta^{*}}$ is the attacker's optimized gradients of the ground-truth gradients $\vec{\nabla\Theta}$. { The Frobenius norm of a matrix can be expressed as $\|\vec{A}\|_F^2 = \sum_{i=1}^{r} \sigma_i^2(\vec{A})$, where $\sigma_i(\cdot)$ denotes the $i$-th singular value of $\vec{A}$ and $r$ is the rank of $\vec{A}$.}

{
By applying truncated SVD (tSVD) on $\vec{\nabla\Theta}$ to retain the top-$k$ singular values based on the energy threshold $\mathcal{T}$, we have:
\begin{equation}
\label{equ:tt}
\sum_{i=1}^{k} \sigma_i^2(\vec{\nabla\Theta}) \geq \mathcal{T} \cdot \|\vec{\nabla\Theta}\|_F^2.
\end{equation}
With tSVD, the attacker observes only the truncated gradients. Therefore, the error between the ground-truth gradients and optimized gradients becomes:
\begin{equation}
\begin{aligned}
    & \mathbb{E} \|\vec{\nabla\Theta} -\vec{\nabla\Theta^{*}} \|_F \\
    & = \mathbb{E} \|\vec{\nabla\Theta} - \mathrm{tSVD}(\vec{\nabla\Theta},\mathcal{T}) + \mathrm{tSVD}(\vec{\nabla\Theta},\mathcal{T}) - \vec{\nabla\Theta^{*}} \|_F \\
    & \overset{(a)}{\leq} \| \mathrm{tSVD}(\vec{\nabla\Theta},\mathcal{T}) - \vec{\nabla\Theta^{*}} \|_F + \| \vec{\nabla\Theta} - \mathrm{tSVD}(\vec{\nabla\Theta},\mathcal{T}) \|_F\\
    & \overset{(\ref{equ:def})}{\leq} \varepsilon + \| \vec{\nabla\Theta} - \mathrm{tSVD}(\vec{\nabla\Theta},\mathcal{T}) \|_F \\
    & =  \varepsilon + \sqrt{ \sum_{i=k+1}^{r} \frac{ \sigma_i(\vec{\nabla \Theta})^{2} } {\sum_{j=1}^r \sigma_j(\vec{\nabla \Theta})^{2}} } \|\vec{\nabla\Theta}\|_F \\
    & \overset{(\ref{equ:tt})}{\leq} \varepsilon +  \sqrt{(1 -\mathcal{T})}\|\vec{\nabla\Theta}\|_F \\
    & = \varepsilon + \sqrt{\gamma_1} \|\vec{\nabla\Theta}\|_F,
\end{aligned}
\end{equation}
where $\gamma_1 = 1-\mathcal{T}$. (a) is based on the Frobenius norm triangle inequality, which states that for any two matrices A and B, the inequality $\|A + B\|_F \leq \|A\|_F + \|B\|_F$ holds.

By augmenting tSVD with Channel-Wise Weighted Approximation, the error between the ground-truth gradients and optimized gradients becomes:
\begin{equation}
\begin{aligned}
    & \mathbb{E} \|\vec{\nabla\Theta} -\vec{\nabla\Theta^{*}} \|_F \\
    & = \mathbb{E} \|\vec{\nabla\Theta} - \vec{I}^{-1}\mathrm{tSVD}(\vec{I}\vec{\nabla\Theta},\mathcal{T}) \\
    & ~~~~ + \vec{I}^{-1}\mathrm{tSVD}  (\vec{I}\vec{\nabla\Theta},\mathcal{T}) -\vec{\nabla\Theta^{*}} \|_F \\
    & \overset{(a)}{\leq} \| \vec{I}^{-1}\mathrm{tSVD}  (\vec{I}\vec{\nabla\Theta},\mathcal{T}) -\vec{\nabla\Theta^{*}} \|_F \\
    & ~~~~ + \| \vec{\nabla\Theta} - \vec{I}^{-1}\mathrm{tSVD}(\vec{I}\vec{\nabla\Theta},\mathcal{T}) \|_F \\
    & \overset{(\ref{equ:def})}{\leq} \varepsilon + \| \vec{\nabla\Theta} - \vec{I}^{-1}\mathrm{tSVD}(\vec{I}\vec{\nabla\Theta},\mathcal{T}) \|_F \\
    & = \varepsilon + \|\vec{I}^{-1}(\vec{I}\vec{\nabla\Theta} - \mathrm{tSVD}(\vec{I}\vec{\nabla\Theta},\mathcal{T}))\|_{F} \\
    & \overset{(b)}{\leq} \varepsilon + \|\vec{I}^{-1}\|_{2} \|\vec{I} \vec{\nabla\Theta} - \mathrm{tSVD}(\vec{I}\vec{\nabla\Theta},\mathcal{T})\|_{F} \\
    & \overset{(\ref{equ:tt})}{\leq} \varepsilon + \sigma_{max}  (\vec{I}^{-1})\sqrt{(1-\mathcal{T})}\|\vec{I}\vec{\nabla\Theta}\|_F \\
    & \overset{(b)}{\leq} \varepsilon +  \sigma_{max} (\vec{I}^{-1})\sqrt{(1-\mathcal{T})}\|\vec{I}\|_{2}\| \vec{\nabla\Theta}\|_F \\
    & = \varepsilon +  \sigma_{max}(\vec{I}^{-1})  \sigma_{max} (\vec{I})\sqrt{(1-\mathcal{T})}\|\vec{\nabla \Theta}\|_F  \\
    & = \varepsilon + \frac{\sigma_{max}(\vec{I})}{\sigma_{min}(\vec{I})} \sqrt{(1-\mathcal{T})}\|\vec{\nabla \Theta}\|_F  \\
    & = \varepsilon + \sqrt{\gamma_2} \|\vec{\nabla\Theta}\|_F,
\end{aligned}
\end{equation}
where $\|\cdot\|_2$ represents the spectral norm, i.e., the maximum singular value of the matrix, and $\gamma_2 = (\frac{\sigma_{max}(\vec{I})}{\sigma_{min}(\vec{I})})^2 (1-\mathcal{T})$. (b) is based on the submultiplicativity of matrix norms, which states that for any matrices A and B, the inequality $\|AB\|_F \leq \|A\|_2 \cdot \|B\|_F$ holds.
Since $\frac{\sigma_{max}(\vec{I})}{\sigma_{min}(\vec{I})} \geq 1$, we have $\gamma_2 \geq \gamma_1$ under the same energy threshold $\mathcal{T}$. This indicates that tSVD with Channel-Wise Weighted Approximation provides stronger protection than the original tSVD. Hence, this theorem holds.
}

\subsection{Visualization of our FL Testbed and Reconstructed Examples}
\label{app:visualization-reconstructed}

{
Fig.~\ref{system} illustrates our FL testbed.
To visualize defense effectiveness, we present examples of reconstructed CIFAR-10 and EMNIST images under different defense methods in Fig.~\ref{fig:CIFAR10_visual} and Fig.~\ref{fig:MNIST_visual}, respectively. We can observe that the reconstructed examples from \textit{SVDefense} are not recognizable for both datasets.
}

\subsection{ARTIFACT APPENDIX}
\subsubsection{Description \& Requirements}
We provide \textit{SVDefense}, a novel defense framework against GIAs that leverages the truncated Singular Value Decomposition (SVD) to obfuscate gradient updates. \textit{SVDefense} integrates three core functionalities: \textit{SVDefense} introduces three key innovations, the Self-Adaptive Energy Threshold that adapts the privacy protection for clients with different vulnerability to GIAs caused by their varying degrees of class imbalance, the Channel-Wise Weighted Approximation that selectively preserves essential gradient information for model training while enhancing privacy protection, and the Layer-Wise Weighted Aggregation for effective aggregation of client updates under class imbalance. We have developed a comprehensive platform that evaluate the defense performance and classification performance of existing defenses and \textit{SVDefense}. In this artifact, we take the cifar10 dataset as an example.

\begin{itemize}
    \item \textbf{How to access}: Our implementation is available on
Zenodo with DOI: https://doi.org/10.5281/zenodo.16948135.
    \item \textbf{Hardware dependencies}: GPU: 8GB (optional), CPU: 8 cores, RAM: 16GB and Disk space: 100 GB of space.
    \item \textbf{Software dependencies}: 
    \begin{enumerate}
        \item Operation System: Ubuntu 22.04.
        \item Package management system: Conda (or Miniconda).
    \end{enumerate}
    \item \textbf{Benchmarks}: cifar10.
\end{itemize}

\subsubsection{Artifact Installation \& Configuration}
\begin{itemize}
    \item \textbf{Create workspace}: Download the code repository from GitHub, and
name as \texttt{˜/svdefense} workspace.
    \item \textbf{Environment setup}: Navigate to \texttt{˜/svdefense/scripts} and run the setup script \texttt{./setup.sh} to configure the development environment.
    \item \textbf{Software}: Install the parallel using \texttt{sudo apt-get install parallel} to run experiments in parallel.
\end{itemize}



\subsubsection{Experiment E1: Defense performance under IG attack} about [20 human-minutes + 100 compute-hours] This experiment aims to:
\begin{itemize}
    \item Assess the functionality of the \textit{Svdefense}.
    \item Evaluate \textit{Svdefense}’s effectiveness under the IG attack, as Table III details.
\end{itemize}

\textbf{[Preparation]}
\begin{itemize}
    \item Change the \texttt{defense} parameter in the \texttt{˜/svdefense/scripts/defense\_IG.sh} to evaluate different defense methods. Here we provide some examples including `none', `dp', 'outpost', `prune', `dgp', `pfgd', and `svdefense'. We can comment out the command to test the performance of the defense.
    \item Using the `parallel' command can reconstruct multiple images concurrently. The usage of the `parallel' command can be found in the \texttt{defense\_IG.sh}. The default command is to reconstruct one image for testing. 
\end{itemize}

\textbf{[Execution]} 
\begin{itemize}
\item Main Script: Navigate to \texttt{˜/svdefense/scripts} and run the setup script \texttt{./defense\_IG.sh}.
\end{itemize}

\textbf{[Results]}
The reconstructed images can be found in the folder \texttt{˜/svdefense/IG\_attack/recon\_data/}, and the ground-truth images can be found in the \texttt{˜/svdefense/IG\_attack/gt\_data/}. Then we can use the \texttt{cal\_matric.py} to output the metrics in Table III.

\subsubsection{Experiment E2: Training perturbation-based defense performance under IG attack} about [2 human-minutes + 1 compute-hours] This experiment aims to:
\begin{itemize}
    \item Evaluate existing training perturbation-based defenses’ effectiveness under the adaptive attack, as Table III details.
    \item We take the PRECODE as an example.
\end{itemize}

\textbf{[Preparation]}
\begin{itemize}
    \item Comment out the code in the 137 line of \texttt{PRECODE/invertinggradients/inversefed /reconstruction\_algorithms.py} to enable the adaptive attack and vice versa.
\end{itemize}

\textbf{[Execution]} 
\begin{itemize}
\item Main Script: Navigate to \texttt{˜/svdefense/scripts} and run the setup script \texttt{./defense\_PRECODE.sh}.
\end{itemize}

\textbf{[Results]}
The reconstructed images can be found in the folder \texttt{˜/svdefense/PRECODE/recon\_data/}, and the ground-truth images can be found in the \texttt{˜/svdefense/PRECODE/gt\_data/}. Then \texttt{cal\_matric.py} can output the metrics in Table III

\subsubsection{Experiment E3: Defense performance under ROG attack}
\quad

\textbf{[Preparation]}
\begin{itemize}
    \item Download the needed pretrained weights following the public GitHub repository \url{https://github.com/KAI-YUE/rog}.
    \begin{enumerate}
        \item Download the pretrained models \footnote{https://huggingface.co/erickyue/rog\_modelzoo/tree/main} and put them under \texttt{ROG\_attack/model\_zoos/}.
        \item Download the csv file \footnote{https://storage.googleapis.com/openimages/v6/oidv6-class-descriptions.csv} and put it under \textit{ROG\_attack/data} folder.
    \end{enumerate}
    \item Change the \texttt{defense} parameter in the configuration file \texttt{ROG\_attack/utils/config\_fedavg.yaml} to evaluate different defenses. Here we provide some examples including `none',`dp', 'outpost', `prune', `dgp', `pfgd', and `svdefense'. 
\end{itemize}

\textbf{[Execution]} 
\begin{itemize}
\item Main Script: Navigate to \texttt{˜/svdefense/scripts} and run the script \texttt{./defense\_ROG.sh}.
\end{itemize}

\textbf{[Results]}
The reconstructed images can be found in the folder \texttt{˜/svdefense/ROG\_attack/recon\_data/}, and the ground-truth images can be found in the \texttt{˜/svdefense/ROG\_attack/gt\_data/}. Then \texttt{python cal\_matric.py} can output the metrics in Table V.

\subsubsection{Experiment E4: Classification performance on cifar10 dataset}
\quad

\textbf{[Preparation]}
\begin{itemize}
    \item Here we provide some examples including `none', 'outpost', `prune', `dgp', `pfgd', and `svdefense'. We can update the parameters in \texttt{˜/svdefense/svd-defense/pyproject.toml} to test the performance of the defense. Noted that, to control for confounding factors, the layer-wise weighted aggregation is omitted so that the analysis focuses solely on the defense applied to gradients.
    \item We can change the `{local-defense}' parameter in the configuration file to evaluate different defense methods in the Federated learning setting.
\end{itemize}

\textbf{[Execution]} 
\begin{itemize}
\item Main Script: Navigate to \texttt{˜/svdefense/scripts} and run the script \texttt{./fl.sh}.
\item Change the parameters `server-device' and `client-device' to `cuda' to accelerate the training process.
\end{itemize}

\textbf{[Results]}
The accuracy across epochs can be found in \texttt{˜/svdefense/svd-defense/\{defense\}\_acc.txt}. Then \texttt{python draw.py} can output one line of results of the corresponding defense in Fig.7.

\end{document}